\documentclass[12pt]{article}

\usepackage[english]{babel}
\usepackage[margin=1in]{geometry}

\usepackage{amsmath}
\usepackage{amsfonts}
\usepackage{bbm}
\usepackage{graphicx}
\usepackage{subcaption}
\usepackage[colorlinks=true, allcolors=blue]{hyperref}
\usepackage[nonamebreak]{natbib}
\usepackage[title]{appendix}
\usepackage[dvipsnames]{xcolor}
\usepackage{algorithm}
\usepackage{algorithmic}
\usepackage{xr-hyper} % use xr-hyper if using hyperref
\newcommand{\E}{\mathbbm{E}}
\newcommand{\mI}{\boldsymbol{\mathcal{I}}}
\newcommand{\bA}{\boldsymbol{A}}

\newcommand{\bU}{\boldsymbol{U}}
\newcommand{\bo}{\boldsymbol{0}}
\newcommand{\bx}{\boldsymbol{X}}
\newcommand{\bz}{\boldsymbol{Z}}
\newcommand{\bi}{\boldsymbol{i}}
\newcommand{\blam}{\boldsymbol{\lambda}}
\newcommand{\bbe}{\boldsymbol{\beta}}

\newcommand{\var}{\text{Var}}
\newcommand{\cov}{\text{cov}}
\newcommand{\what}{\widehat} 
\newcommand{\One}{\mathbbm{1}} 
\newcommand{\tRsq}{R^2}
\newcommand{\cIbl}{\mI_{\bbe | \blam}}
\newcommand{\tef}{f^2} 

\DeclareMathOperator*{\argmin}{arg\,min}

\title{\textbf{Robust Power and Sample Size Calculations in Quasi-likelihood Models: Methods and Practice}}
\author{You}
\author{
Shijie Yuan\thanks{Department of Statistics and Data Sciences, The University of Texas at Austin, Austin, TX 78712, USA; Corresponding Email: sj.yuan@austin.utexas.edu}\ ,
Amy Cochran\thanks{Department of Math, Department of Population Health Sciences, University of Wisconsin Madison, Madison, WI 53706, USA; Email: cochran4@wisc.edu}\ ,
and 
Paul Rathouz\thanks{Department of Population Health, The University of Texas at Austin, Austin, TX 78712, USA; Email: paul.rathouz@austin.utexas.edu}\ 
 	}

\begin{document}
\maketitle
\thispagestyle{empty}

\newpage
\thispagestyle{empty}
\begin{abstract}
\noindent \textbf{Background}:
Accurate power and sample size calculations are essential in study planning, yet they are often difficult to carry out for quasi-likelihood (QL) models. Traditional PSS approaches often rely on restrictive distributional assumptions, 
limiting their applicability when responses have non-standard distributions, variance functions are misspecified, or when predictors exhibit complex dependence structures.

\noindent \textbf{Methods}:
We examine whether two effect size measures—2 Standard Deviations in the Linear Predictor (2SLiP) and Pseudo-Partial $R^2$ (P2R2)—originally developed for Wald tests involving generalized linear models, are effective at power and sample size calculations in the QL framework. Through extensive simulations across diverse outcome types, link functions, and variance structures, we assess their performance under Wald tests and explore whether they remain useful for score tests.
To illustrate practical utility, we apply these effect size measures to survey data on frontline health care workers to quantify the association between perceived personal protective equipment (PPE) adequacy and burnout risk during the COVID-19 pandemic, adjusting for covariates.

\noindent \textbf{Results}: We show that the two GLM-based effect sizes are fundamentally moment-based objects, and therefore extend directly to QL models. Across all simulation settings, both measures remained accurate, with sample size and power estimates within 3\% and 2\% of the target, respectively. In the case study, the estimated effect sizes for perceived PPE adequacy (2SLiP = 0.096 and P2R2 = 0.020) correspond to small but meaningful associations with burnout risk after adjustment for demographic and occupational covariates; both yielded accurate recommendations for sample size.

\noindent \textbf{Conclusions}:
Both 2SLiP and P2R2 offer robust alternatives for power and sample size calculation in QL settings. By requiring minimal distributional assumptions, these measures enhance the flexibility and reliability of power and sample size calculations for realistic study designs commonly encountered in medical and public health research.
\end{abstract}

% We show that GLM-based effect sizes are fundamentally moment-based objects, and therefore extend directly to quasi-likelihood models, enabling power calculations in a far broader and more robust class of estimating-equation settings.


\clearpage\newpage
\setcounter{page}{1}
\section{Introduction}

Modern research increasingly relies on quasi-likelihood (QL) models to analyze outcomes ranging from binary events and counts to continuous measurements. Unlike generalized linear models (GLMs), which require full distributional specification within the exponential family, QL methods only posit a mean–variance relationship and a link function, allowing valid inference about the mean model under weaker assumptions \citep{wedderburn1974quasi}. This flexibility makes QL models especially attractive in applied research, where outcomes often deviate from canonical distributions, predictors exhibit complex correlation structures, and overdispersion is common. Yet, the same generality complicates study design: there remains a lack of generalized effect size measures that can accommodate arbitrary numbers of predictors and adjustors as well as diverse outcome types, leaving power and sample size (PSS) estimation dependent on bespoke solutions for each new problem and setting. 

Over the past 40 years, a number of advances have been made within the framework of GLMs. Some, but not all, of these results might generalize to QL models, even when such extensions are not made explicit by the original authors. Several methods have specifically focused on the Wald test with a single predictor.
Hauck and Donner introduced the use of the Wald test in logistic regression for testing hypotheses on the coefficient of a single predictor, deriving the Fisher information and the conditional variance of the estimator \citep{hauck1977wald}.
Whittemore developed PSS approximations for logistic regression with rare outcomes, approximating the Fisher information through the moment-generating function of the covariates \citep{whittemore1981sample}.
% and thereby facilitating inference in generalized linear models with small response probabilities.
Signorini later extended this idea to the Poisson regression model, providing exact, albeit asymptotic,  calculations without relying on approximation \citep{signorini1991sample}. 
Wilson and Gordon proposed a general framework for PSS determination in GLMs that accounts for confounding variables by deriving approximate coefficient variances under both null and alternative hypotheses \citep{wilson1986calculating}.
% —explicitly recognizing that these variances differ due to the change in the maximum likelihood estimates of the confounding variables across hypotheses. 
Building on this, Shieh refined Wald-based sample size methods by adjusting the variance of the tested coefficient according to the limiting values of nuisance parameters under the null hypothesis \citep{shieh2001sample}, extending earlier work on score and likelihood ratio tests \citep{self1988power,self1992power}.

Subsequent studies expanded Wald-based approaches to multiple predictors \citep{shieh2005power,lyles2007practical}, with particular emphasis on nominal, count, and ordinal outcomes \citep{lyles2007practical}. 
% Both citep{shieh2005power} and citep{lyles2007practical} extended Wald test-based power and sample size methods to handle multiple predictors in GLMs, although citep{lyles2007practical} focuses on nominal, count, and ordinal outcomes.
Other contributions have addressed GLM-based PSS \citep{hsieh1998simple,demidenko2007sample,demidenko2008sample,novikov2010modified,channouf2014power,bush2015sample}, though most remain limited to logistic or Poisson regression and specific covariate settings.
Common software (e.g., SAS PROC POWER, G*Power, PASS) reflects these restrictions, focusing on logistic, Poisson, or ANOVA-style designs \citep{o1984procedures,faul2007g,hintze2006pass}.
To address such limitations, Igeta proposed a QL–based framework for PSS calculation that explicitly incorporates possible misspecifications of the variance function in trials with over-dispersed count data \citep{igeta2018power}. Beyond assumptions about the response distribution, traditional approaches also impose strong and often complex requirements on the joint distribution of predictors and adjustors, as well as on the exact outcome–predictor relationships. These demands limit their practical use, especially with multiple predictors or complex study designs \citep{demidenko2007sample, novikov2010modified}, such as when covariates arise as random processes in observational studies. As a result, researchers are frequently forced to adopt unrealistic assumptions or oversimplified models, reducing both the reliability of findings and the efficiency of resource use.

Addressing these limitations, Cochran et al. proposed two effect size measures for PSS calculations applicable to any GLM \citep{cochran2025glm}. For ease of reference, we refer to these as \underline{2} \underline{S}tandard deviations in \underline{Li}near \underline{P}redictor (2SLiP), which captures the additional variance explained by predictors in the linear predictor scale beyond adjustors, and Pseudo-Partial $R^2$ (P2R2), which generalizes partial $R^2$ to GLMs by quantifying the proportion of outcome variance on a standard deviation scale explained by predictors beyond adjustors. These measures were designed to benefit from the same properties that make the framework of citep{gatsonis1989multiple} useful for study planning in linear regression, namely, interpretability, support for joint testing of predictors, adjustment for arbitrary covariates, and reliance only on first and second moment information.  Because the development of these effects sizes is restricted to GLMs, this restriction implicitly ties the proposed effect size measures to fully specified likelihoods. By exploiting the fact that 2SLiP and P2R2 depend only on first- and second-moment information, we extend the framework to QL models, broadening its applicability while preserving interpretability and accessibility.

This article extends prior methodological work in four ways:  First, we launch from a QL framework, a semi-parametric generalization of GLMs that requires only specification of the linear predictor, link, and variance function, without a full probability model \citep{mccullagh1989generalized}. Second, we
examine whether the proposed PSS calculations remain useful when applied to score tests, moving beyond the Wald test focus of earlier work. Third, we evaluate the performance of 2SLiP and P2R2 through simulations across diverse QL settings, including overdispersed Poisson, dispersed gamma, and non-canonical links, to assess accuracy and robustness under realistic data structures. Fourth, we present a case study using survey data from citep{cahill2022occupational} to demonstrate how these measures can be applied in practice, linking effect size estimates to sample size requirements in studies of perceived personal protective equipment (PPE) adequacy and mental health.

The remainder of this article is organized as follows. Section \ref{sec:methods} reviews the proposed effect size measures, 2SLiP and P2R2 and extends the methodological framework to the QL models and score tests, and reviews the proposed effect size measures, 2SLiP and P2R2. Section \ref{sec:sim} reports simulation studies evaluating their performance across cases. Section \ref{sec:case} presents a case study illustrating the application of these measures in practice. Section \ref{sec:dis} discusses results, practical implications, and directions for future research.

\section{Methods}\label{sec:methods}

\subsection{Preliminaries}

Consider an outcome \( Y \) for a single unit together with its predictors \( \bx \) and  adjustors \( \bz \). The vector of predictors \( \bx \) is \( p \)-dimensional, while \( \bz \) is \( r \)-dimensional capturing additional covariates (adjustors), including an intercept term.

Under QL theory \citep{mccullagh1983quasi}, the user specifies the mean and variance of $Y$ conditional on $\bx$ and $\bz$ as 
$$\E[Y \mid \bx, \bz] = \mu(\bx,\bz) \qquad \text{and} \qquad \var(Y \mid \bx, \bz) = \sigma^2 v(\mu(\bx,\bz)),$$ 
where $\sigma^2$ is a dispersion parameter and $v(\cdot)$ is a known variance function. We refer to $\mu(\bx,\bz)$ simply by $\mu$ and $v(\mu(\bx,\bz))$ by $v$ when unambiguous. The mean $\mu$ is related to the predictors and adjustors through a known link function $g(\mu)  = \blam' \bz + \bbe' \bx  \equiv \eta$, where $\eta$ is the linear predictor, and $\blam$ and $\bbe$ are the parameter vectors. 

 Parameters $\blam$ and $\bbe$ are estimated by solving the quasi-score equations. These equations are recovered from the quasi-score function, given by
\begin{equation}\label{eq:score_f}
    \bU(\blam, \bbe) =  \frac{Y - \mu}{\sigma^2 v(\mu)} \left(\frac{d\mu}{d\eta} \right) \begin{bmatrix} \bz \\ \bx \end{bmatrix}.
\end{equation}
In practice, parameter estimates are obtained by setting the sample mean of the quasi-score function to zero. In the population, this corresponds to solving $\E[\bU(\blam, \bbe)] = 0$ with expectation taken over the true joint distribution of $(Y,\bx,\bz)$. Solutions to this population equation, denoted by \( (\blam^*, \bbe^*) \), define the population QL parameter values.

Directly solving the population quasi-score equations is generally infeasible due to the nonlinearity of the link function. However, an equivalent characterization is provided by iteratively weighted least squares (IRLS) \citep{mccullagh1989generalized}. 
Specifically, by taking a first-order Taylor expansion of the outcome $Y$ transformed to the linear predictor scale and expanded about its mean, we obtain a working linearized outcome for each observation: \(Y_l \equiv \eta + \frac{d\eta}{d\mu}(Y-\mu).\)
The quasi-score equations can then be replaced by normal equations of a weighted least squares problem, in which $Y_l$ is repeatedly regressed onto $\eta$ with weights for a single observation given by
\begin{equation}\label{eq:weight}
w
\equiv
\frac{1}{\sigma^2 v(\mu)}
\left(\frac{d\mu}{d\eta}\right)^2.
\end{equation}
This can be equivalently expressed with the first-order conditions for minimizing a weighted mean squared error:
\begin{equation}\label{eq:min_wmse}
(\blam^*,\bbe^*)
=
\argmin_{\blam,\bbe}
\E\left[
w
\left\{
Y_l - \big(\blam' \bz + \bbe' \bx\big)
\right\}^2
\right].
\end{equation}
%Moving forward, all quantities evaluated at the quasi-true parameter values $(\blam^*,\bbe^*)$.

\subsection{Measures of effect size}

The key observation of the present paper is that the weighted mean squared error in Equation~\eqref{eq:min_wmse} has exactly the same form as in the GLM setting: it depends only on the conditional mean of $Y$ given $(\bx,\bz)$, as specified by the link function and linear predictor, and on the conditional variance of $Y$. In GLMs, this variance is determined by the assumed distributional family, whereas under QL it is specified directly as $\var(Y\mid\bx,\bz)=\sigma^2 v(\mu)$, without otherwise specifying the distribution of $Y$. Consequently, any construction that depends only on the weighted least squares representation---and not on a fully specified likelihood---generalizes from the GLM to the QL setting.

In particular, the effect size measures introduced for GLMs in citep{cochran2025glm} are of this form, depending only on the weighted least squares representation. We therefore adopt the same constructions here, with expectations and weights taken under the QL specification. To formalize these measures in the current setting, we first introduce the reduced model that includes only the adjustors $\bz$. Let
\begin{equation}\label{eq:min_wmse_reduced}
\blam^*_0
=
\argmin_{\blam}
\E\!\left[
w
\left\{
Y_l - \blam' \bz
\right\}^2
\right],
\end{equation}
where $w$ and $Y_l$ are evaluated as in \eqref{eq:min_wmse} at the true QL parameter values $(\blam^*,\bbe^*)$. The reduced-model linear predictor and conditional mean are then given by
\[
\eta_z = {\blam^*_0}' \bz,
\qquad
\mu_z = g^{-1}(\eta_z).
\]

Using these quantities, we define two effect size measures. The first is a \underline{$2$} \underline{S}tandard deviation contrast on the \underline{Li}near \underline{P}redictor scale (2SLiP),
\begin{equation}\label{eq:2slip}
\phi_{x\mid z}
\equiv
2\,\sqrt{\var(\eta_{x|z})},
\end{equation}
where $\eta_{x|z} \equiv \eta - \eta_z$. With $\eta_{x|z}$ representing the difference between the full linear predictor $\eta$ and the reduced-model linear predictor $\eta_z$ that includes only $\bz$, 2SLiP quantifies the variability in the linear predictor attributable to $\bx$ beyond $\bz$. The second measure is a \underline{P}seudo-\underline{P}artial \underline{$R$}-squared (P2R2),
\begin{equation}\label{eq:p2r2}
\tRsq_{x\mid z}
\equiv
\frac{\E\left[(Y - \mu_z)^2 / \{\sigma^2 v(\mu)\}\right] - \E\!\left[(Y - \mu)^2 / \{\sigma^2 v(\mu)\}\right]}
{\E\left[(Y - \mu_z)^2 / \{\sigma^2 v(\mu)\}\right]}.
%\frac{\E\!\left[(\mu - \mu_z)^2 / \{\sigma^2 v(\mu)\}\right]}
%{1 + \E\!\left[(\mu - \mu_z)^2 / \{\sigma^2 v(\mu)\}\right]}.
\end{equation}
It measures the proportion of outcome variability on a standard deviation scale explained by the full model mean $\mu$ beyond the reduced-model mean $\mu_z$ that includes only $\bz$.  In the following section, we outline how these effect size measures are employed to obtain approximate power calculations, and we provide theoretical justification for these approximations in Supplementary Section {\color{blue} S.2}. 

Because both effect size measures depend only on the first two conditional moments of Y given $(\bx,\bz)$, their interpretation can be anchored in the GLM setting and carried over. For example, in Poisson regression with a log link, 2SLiP corresponds to the log rate ratio for $Y$ contrasting two units that differ by two standard deviations in $\eta_{x|z}$; for a balanced binary predictor $\bx$ with no adjustors, this is simply the contrast between $\bx=1$ and $\bx=0$. In linear regression with unit error variance, 2SLiP reduces to Cohen's d \citep{cohen1992power}, and P2R2 is exactly the partial $R^2$. The only distinction in the QL setting is that the conditional variance, and hence the weights, may include a dispersion parameter $\sigma^2$, which affects the scaling of variability but not the interpretation of the effect size. In an overdispersed Poisson quasi-likelihood model, for instance, 2SLiP retains the interpretation of a log rate ratio over a two–standard deviation contrast in $\eta_{x|z}$.

\subsection{Power calculations}

The effect size measures introduced are intended to support power and sample size calculations across a broad class of models, including those not fully specified by a parametric family. In GLMs, these measures can be used directly in Wald test power calculations. We ask whether this approach is also appropriate for QL models, considering both Wald and score tests evaluating the null hypothesis that the predictors of interest have no effect, i.e. $H_0:\bbe^* = \bo$. We review the relevant tests and describe how the effect sizes are applied.

\paragraph{Wald test.} In QL models with $n$ independent samples, the Wald statistic is constructed from the QL estimators of the model parameters and their estimated covariance, denoted by $(\what{\blam}, \what{\bbe})$ and $\bi_n^{-1}(\what{\blam}, \what{\bbe})$, respectively. The matrix $\bi_n(\blam, \bbe)$ is the expected Fisher information matrix in QL theory and can be computed from the quasi-score function $\bU_i(\lambda,\beta)$ evaluated at  i.i.d. observations $(Y_i, \bx_i, \bz_i)$ for $i=1$ to $i=n$. Compactly, we can write this matrix as
\begin{equation}\label{eq:information}
    \bi_n(\blam,\bbe) = \sum_{i=1}^n \E\left[ - \frac{\partial\bU_i(\blam, \bbe)}{\partial(\blam, \bbe)} \bigg|  \bx_i, \bz_i\right] = \sum_{i = 1}^n w_i \begin{bmatrix} \bz_i\bz_i' & \bz_i\bx_i'\\ \bx_i\bz_i' & \bx_i\bx_i' \end{bmatrix},
\end{equation}
whose dependence on $\blam$ and $\bbe$ enters only through $w_i$, representing the weight term evaluated at the observation $(Y_i, \bx_i, \bz_i)$. That is, $w_i = (\frac{d\mu}{d\eta} |_{\eta = \eta_i})^2 / \left( \sigma^2 v(\mu_i) \right)$ with $\eta_i = \blam'\bz_i + \bbe'\bx_i$ and $\mu_i = g^{-1}(\eta_i)$. 
Taking $\bi_{n,\bbe|\blam}^{-1}(\blam,\bbe)$ to be the lower right $p \times p$ block of $\bi_{n}^{-1}(\blam,\bbe)$, the Wald test statistic becomes $\what{W} =  \what{\bbe}' \bi_{n,\bbe|\blam}(\what{\blam}, \what{\bbe}) \what{\bbe}.$

Asymptotically, the Wald statistic $\what{W}$ follows a non-central chi-squared distribution $\chi^2_p(\Delta)$ with \(p\) degrees of freedom and non-centrality parameter $\Delta \equiv n \tef$ \citep{mccullagh1983quasi}.  A formal derivation of the asymptotic distribution of the Wald statistic in the quasi-likelihood setting is provided in Supplementary Section {\color{blue} S.1}.  The effect size measure $\tef$ is defined as
\begin{equation} \label{eq:true_effect}
    \tef \equiv {\bbe^*}' \cIbl \bbe^*,
\end{equation}
where
\begin{equation}\label{eq:cond_fisher_info}
    \cIbl \equiv \E\left[ w \bx \bx' \right] - \E\left[ w \bx \bz' \right] \E\left[ w \bz \bz' \right]^{-1} \E\left[ w \bz \bx' \right],
\end{equation}
obtained as the Schur complement of the block $\E[w \bz \bz']$ in the marginal Fisher information matrix:
\begin{align} \label{eq:pop_fisher}
\mI =
\begin{bmatrix}
\E[w \bz \bz'] & \E[w \bz \bx'] \\
\E[w \bx \bz'] & \E[w \bx \bx']
\end{bmatrix}.
\end{align}
Note $\cIbl^{-1}$ corresponds to the lower right $p \times p$ block of $\mI^{-1}$.

Once you have the effect size $f^2$, calculating the asymptotic power of the Wald test is straightforward. Since the Wald statistic follows a central chi-squared distribution under the null hypothesis, you would reject at significance level $\alpha$ if $\what{W}$ exceeds the $(1-\alpha)$-quantile of $\chi^2_p(0)$. Under the alternative hypothesis, the asymptotic power is then simply the probability that a non-central chi-squared distribution with non-centrality parameter $\Delta = n f^2$ exceeds the same critical value used under the null. Formally, the asymptotic power is given by
\begin{equation} \label{eq:power}
    q = 1 - F_{\chi^2_p(\Delta)} \left( F^{-1}_{\chi^2_p(0)}(1 - \alpha) \right).
\end{equation} 
is the cumulative distribution function of the non-central chi-squared distribution. In other words, once you have $f^2$ and the sample size $n$, you can immediately compute asymptotic power for any chosen significance threshold.

\paragraph{Approximate power.} Because $f^2$ depends on population quantities that are typically unknown, we approximate it using the effect size measures 2SLiP \eqref{eq:2slip} and P2R2 \eqref{eq:p2r2}, and substitute these approximations into the power expression above.  Formal error analyses are available in the GLM setting \citep{cochran2025glm}, and because these arguments depend only on first and second moments, they apply equally to the present QL framework. Below, we present the intuition behind why these approximations work in this broader context.

Both measures are grounded in the identity $f^2 = \E[ w \eta_{x|z}^2 ]$,  with theoretical justification provided  in Supplementary Section {\color{blue} S.2}.  For 2SLiP, expanding this expectation using basic moment identities yields
\begin{equation}\label{eq:true_zeta_decomp}
\tef = \E[w] \left(\E[\eta_{x|z}]^2 + \phi_{x|z}^2 / 4\right) + \cov\left[w, \eta_{x|z}^2\right].
\end{equation}
This decomposition clarifies when a simple approximation is reasonable. In settings where the weights exhibit little variation relative to $\phi_{x|z}$, the mean contribution $\E[\eta_{x|z}]$ is small, the covariance term $\cov[w,\eta_{x|z}^2]$ is negligible, and $\E[w]$ is well approximated by a constant $w_{\One}$. Under these conditions, the effect size $f^2$ is well approximated by
\begin{align*}
f^2 \approx f^2_{\phi}
\equiv w_{\One} \phi_{x|z}^2 / 4.
\end{align*}
This approximation is exact when weights are constant, as in linear regression. In practice, we set $w_{\One}$ to the weight evaluated at $\mu = \E[Y]$, since the marginal mean can often be specified. 

For P2R2, we proceed slightly differently. A first-order Taylor expansion of the conditional mean around $\eta_z$ yields 
\begin{equation}\label{eq:mu_eta}
    \frac{d\mu}{d\eta} (\eta-\eta_z) = \frac{d\mu}{d\eta} \eta_{x|z} \approx (\mu - \mu_z).
\end{equation}
Substituting this approximation into the identity $f^2 = \E[w\,\eta_{x|z}^2]$ gives
\begin{align*}
    f^2 \approx f^2_R \equiv \E\left[ \frac{ (\mu - \mu_z)^2}{\sigma^2 v(\mu)} \right] = \frac{\tRsq_{x\mid z}}{1-\tRsq_{x\mid z}}.
\end{align*}
This approximation is exact if the link function $g(\cdot)$ is linear, as with the identity link. For nonlinear mean functions, the approximation is accurate when $\eta_{x|z}$ is close to zero, so that the first-order Taylor expansion about $\eta_z$ provides a good local approximation.

% This approximation provides the connection between the interpretable effect size 2SLiP and the effect size f^2 that drives power calculations.

\paragraph{Score Test.} \label{sec:score_test}
The preceding develop uses 2SLiP and P2R2 to approximate the non-centrality parameter governing the asymptotic power of the Wald test. While this provides a principled framework, it is ultimately an approximation rooted in large-sample theory, whereas power in practice is determined by finite-sample behavior. This motivates examining whether the same effect-size–based approximation strategy extends is a good heuristic for other testing procedures. To investigate this, we review the score test in the QL setting.

Let $\bU_n(\blam,\bbe)$ denote the function obtained by summing the individual quasi-scores $\bU_i(\blam,\bbe)$ over $n$ i.i.d. observations \((Y_i,\bx_i,\bz_i)\), and let $\bU_{n\blam}(\blam,\bbe)$ represents the part of the vector $\bU_n(\blam,\bbe)$ corresponding to $\blam$. The score test is based on the restricted estimator $\what{\blam}^{(0)}$, defined as the solution to $\bU_{n\blam}(\blam,\bo)=\bo$. The resulting score statistic is:
\begin{equation}\label{eq:score_test}
    \what{S} = \bU_n\left(\what{\blam}^{(0)}, \bo\right)'\bi_n^{-1}\left(\what{\blam}^{(0)}, \bo\right)\bU_n\left(\what{\blam}^{(0)}, \bo\right).
\end{equation}
Under the null hypothesis $\bbe^*=\bo$, $\what{S}$ converges in distribution to a central chi-squared distribution with $p$ degrees of freedom. Under the alternative $\bbe^*\neq\bo$, however, its limiting distribution is no longer chi-squared, which complicates power calculations.

This complication arises because $\what{\blam}^{(0)}$ is generally not a consistent estimator of $\blam^*$ when $\bbe^*\neq\bo$. Instead, it converges to the solution $\blam^{(0)}$ to the population equation $\E[\bU_{\blam}(\blam,\bo)]=\bo$ where $\bU_{\blam}(\blam,\bo)$ is the part of the quasi-score $\bU(\blam,\bbe)$ corresponding to $\blam$. As a result, determining the asymptotic behavior of the score statistic under the alternative requires characterizing both $\blam^{(0)}$ and the corresponding mean and covariance of $\bU_n(\what{\blam}^{(0)},\bo)$. While these quantities simplify under the null, they do not admit comparably simple expressions under the alternative, making exact power calculations difficult.

%The difficulty stems from $\what{\blam}^{(0)}$ not generally being a consistent estimator of $\blam^*$ when $\bbe^* \neq 0$; instead, it approaches the solution $\blam^{(0)}$ of the population equation $\E[\bU_{\blam}(\blam,\bo)] = \bo,$ where $\bU_{\blam}(\blam,\bbe)$ represents the part of the vector $\bU(\blam,\bbe)$ corresponding to $\blam$. So to do power calculations properly, you have to know what $\blam^{(0)}$ is under the alternative - which is not straightforward to figure out.  Further, when $\bbe = 0$, the asymptotic mean and covariance of $n^{-1} \bU_n(\what{\blam}^{(0)}, \bo)$ simplify neatly to $\bo$ and $\cal I.$ However, when $\bbe^* \neq 0$, the asymptotic mean of $n^{-1} \bU_n(\what{\blam}^{(0)}, \bo)$ is no longer zero, and more importantly, its asymptotic covariance is not the asymptotic limit of the matrix $n^{-1} \bi_n^{-1}\left(\what{\blam}^{(0)}, \bo\right)$ featured in the score function. 

Given this difficulty, we adopt a simple heuristic to help benchmark how our power approximations based on 2SLiP and P2R2 perform for score tests in finite samples. Specifically, we define an approximate effect size for the score test by replacing the random components of the score statistic with their population counterparts:
\begin{equation}\label{eq:score_effect}
    f_s^2 \equiv \E\left[\bU(\blam^{(0)}, \bo)'\right]\E\left[\bi^{-1}(\blam^{(0)}, \bo)\right]\E\left[\bU(\blam^{(0)}, \bo)\right].
\end{equation}
Here, $\E[\bi^{-1}(\blam^{(0)}, \bo)]$ represent the marginal Fisher information, obtained by substituting  $\blam = \blam^{(0)}$ and $\bbe = \bo$ into \eqref{eq:pop_fisher}. As with $f^2$, $f_\phi^2$, and $f_R^2$, we use $f_s^2$ by substituting $n f_s^2$ for the noncentrality parameter in the corresponding asymptotic power calculation. This approximation is motivated by behavior near the null and is intended solely as a benchmark rather than an exact characterization.

\section{Simulations} \label{sec:sim}
 
The simulation study evaluates the robustness of 2SLiP and P2R2 for Wald and score tests under GLM and QL settings, assessing whether these measures continue to provide stable and interpretable PSS estimates when standard likelihood assumptions are relaxed. 
% The simulation study evaluates the performance of 2SLiP and P2R2 for Wald and score tests under varying outcome and predictor conditions, with the goal of assessing how well these measures preserve their theoretical properties, particularly in PSS estimation. 
A key focus of this study is on cases where the outcome variable is either a count or a positive continuous quantity. For count outcomes, we examine different link functions (log and identity) and variance structures: variance equal to the mean, proportional to the mean, or proportional to the squared mean. For positive continuous outcomes, we evaluate cases where the variance is proportional to either the mean or the squared mean under a log-link function.

We vary the specification of predictors and adjustors. 
Each simulated dataset includes an intercept, a uniformly distributed covariate $Z$, and a categorical variable $X$. To capture realistic dependence, we use copula-based methods to introduce controlled correlation $\rho$ between $X$ and $Z$ (see Section {\color{blue} S.4} in the Supplemental Materials for details). This setup allows us to investigate how different dependence patterns between predictors and the adjustor affect estimation and inference.
Across all cases, we use a consistent simulation framework by specifying target type I error rate $\alpha = 0.05$ and power $q = 0.8$. Given $\alpha$, $q$ and degrees of freedom $p$, we compute the true non-centrality parameter $\Delta$ by solving equation \eqref{eq:power}. The link function and linear predictor are:
\begin{equation} \label{eq:sim_link}
    g(\mu) = g(\E[Y \mid Z, D_1, D_2]) = \eta =  \lambda_1 + \lambda_2 Z + \beta_1 D_{1} + \beta_2 D_{2},
\end{equation}
where $X \in \{0,1,2\}$ is represented by two dummy variables \(D_{1}\) and \(D_{2}\).
Let $\blam = (\lambda_1, \lambda_2)'$ and $\bbe = (\beta_1, \beta_2)'$. Consider the null hypothesis \( H_0 : \bbe^* = \bo \). Hence, the degrees of freedom $p = 2$, and the true non-centrality parameter needed to achieve 80\%\ power is  $\Delta \approx 9.63$.

\subsection{Wald Test Results} \label{sec:sim_res_wald}
Given the joint distribution of $Z$ and $(D_1,D_2)$ and the coefficients in \eqref{eq:sim_link}, we compute the true asymptotic effect $\tef$ and the approximations $\tef_{\phi}$ and $\tef_R$. Since these do not admit closed-form expressions, we approximate them via Monte Carlo, by sampling $Z$ and $(D_1,D_2)$ into a large data set and computing the associated $w$ and $\eta_{x|z}$. The required sample sizes are then  $n = \lceil \Delta / \tef \rceil$, $n_{\phi} = \lceil \Delta / \tef_{\phi} \rceil$, and $n_R = \lceil \Delta / \tef_R \rceil$, where $\lceil \cdot \rceil$ denotes the ceiling function.
% We set the coefficients $\lambda_1 = 1, \lambda_2 = 0.15, \beta_1 = 0.1$, and $\beta_2 = 0.25$ to achieve a target sample size of $n = 400$ approximately, ensuring $\alpha = 0.05$ and $q = 0.8$ for detecting the true effect $\tef$  under zero correlation ($\rho = 0$) between $Z$ and $(D_1,D_2)$ in Poisson regression with a log link (see Section~\ref{S-sec:coeff_search} in the Supplemental Materials for details).
Let $N \in \{n, n_{\phi}, n_R\}$ be the chosen sample size. For each  replicate, we generate an i.i.d.\ sample $\{(Y_i, Z_{0i}, D_{1i},D_{2i})\}_{i=1}^N$, where $Z_{0i}$ and $(D_1,D_2)$ follow  Section {\color{blue} S.4} in the Supplemental Materials, and outcome $Y_i$ is then generated based on the specified case. 
In the following, we set $\bz = (1, Z)$, and $\bx=(D_{1}, D_{2})$ for each observation. The Wald statistic $\what{W}$ is computed, and $H_0$ is rejected if $\what{W} > F^{-1}_{\chi^2_p(0)}(1 - \alpha)$. Repeating this procedure 10{,}000 times  yields the empirical type I error rate under the null or the empirical power under the alternative, calculated as the rejection proportion.

\subsubsection{Count Outcomes} \label{sec:count}

We investigate three cases of outcome distributions when the outcome is a counted quantity.
\textbf{First}, $Y_i \sim \text{Poisson}(\mu_i)$, labeled ``var = mean",  which corresponds to a correctly specified GLM. The remaining two cases depart from the Poisson mean–variance relationship and are analyzed under a QL framework. 
\textbf{Second}, $Y_i \sim (1 - B_i) \text{Poi}(\mu_i) + B_i\text{Poi}(L_i)$, labeled ``var $\propto$ mean",  
where \( B_i \sim \text{Bernoulli}(0.5) \) and \( L_i \sim \text{Poi}(\mu_i) \) are mutually independent for each $i$. The outcome $Y_i$ generated conditionally on $(B_i, L_i)$. 
\textbf{Third}, labeled ``var $\propto$ mean$^2$", outcomes follow a modified negative binomial distribution, $Y_i \sim \text{NB}(\mu_i,\nu_i)$ with conditional mean and variance are $\E[Y_i \mid Z_i, D_{1i}, D_{2i}] = \mu_i$ and $\var(Y_i \mid Z_i, D_{1i}, D_{2i}) = \mu_i + \nu_i \mu_i^2$. We choose $\nu_i$ such that $\var(Y_i \mid Z_i, D_{1i}, D_{2i}) = \sigma^2 \mu_i^2$, placing this setting within the class of models with a constant coefficient of variation (see Chapter 8 of citep{mccullagh1989generalized}), where the variance is proportional to the square of the mean. Throughout the case of count outcomes, we fix $\sigma^2 = 2$.  This specification yields a non-standard negative binomial distribution for which the MLE is difficult to obtain in closed form. Instead, the estimation of $(\blam, \bbe)$ and the dispersion parameter $\sigma^2$ follows the procedure in Section {\color{blue} S.6} in the Supplemental Materials. 

\paragraph{Log Link}

With targeted type I error rate $\alpha = 0.05$ and power $q = 0.8$, we set the coefficients $\lambda_1 = 1, \lambda_2 = 0.15, \beta_1 = 0.1$, and $\beta_2 = 0.25$ in this simulation study. We use the log link function, so that $\frac{d\mu}{ d \eta} = \mu = \exp(\eta)$. The coefficients are chosen to achieve a target sample size of $n = 400$ approximately under $\rho = 0$ (see Section {\color{blue} S.5} in the Supplemental Materials for details).
Figure~\ref{fig:count_log} presents the sample sizes $n$, $n_{\phi}$, and $n_{R}$ for three cases of outcome variable under this specification across varying values of the correlation parameter $\rho$, along with the corresponding empirical type I error rates and power, obtained from simulations.
As \( \rho \) increases, both \( \var(\eta_{x|z}) \) and the pseudo-partial R-squared \( \tRsq_{x|z} \) decline, because the unique contribution of \( \bx \) to the outcome \( Y \), beyond that of \( \bz \), diminishes as the dependence between the predictor \( \bx \) and the adjustor \( \bz \) strengthens. 
Consequently, larger sample sizes are needed to maintain nominal power, a pattern consistent across all variance structures.
Nevertheless, the impact of \( \rho \) on the ratios \( n_{\phi}/n \) and \( n_R/n_{s} \) is relatively minor, indicating that both 2SLiP and P2R2 remain robust for relative sample size estimation  under varying dependence between predictors ($\bx$) and adjustors~($\bz$). 

\begin{figure}[!htb] 
    \centering
    \begin{subfigure}{.32\textwidth}
    \centering
    \includegraphics[width=\linewidth]{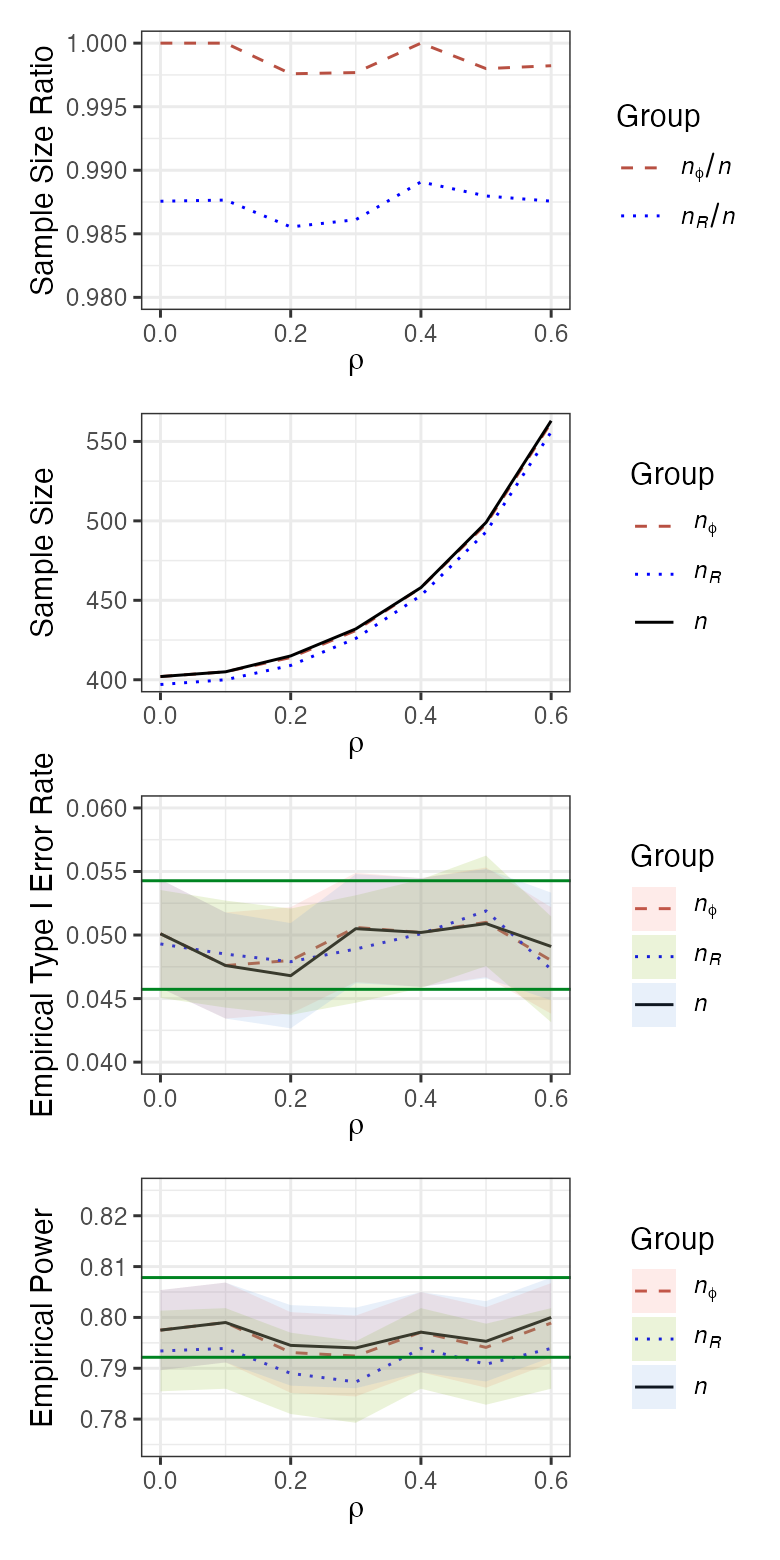}
    \caption{var = mean}
    \end{subfigure}
    \begin{subfigure}{.32\textwidth}
    \centering
    \includegraphics[width=\linewidth]{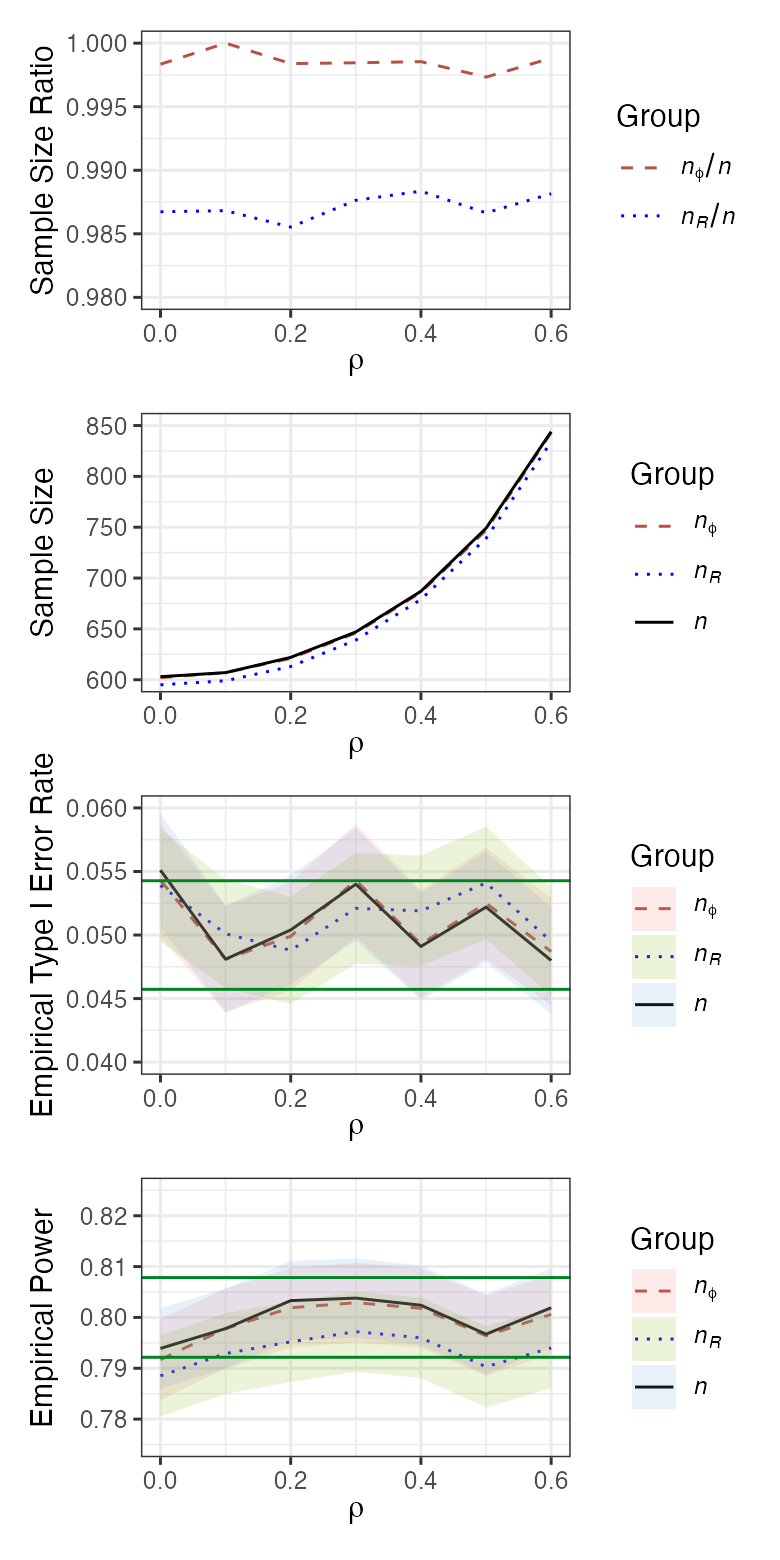}
    \caption{var $\propto$ mean}
    \end{subfigure}
    \begin{subfigure}{.32\textwidth}
    \centering
    \includegraphics[width=\linewidth]{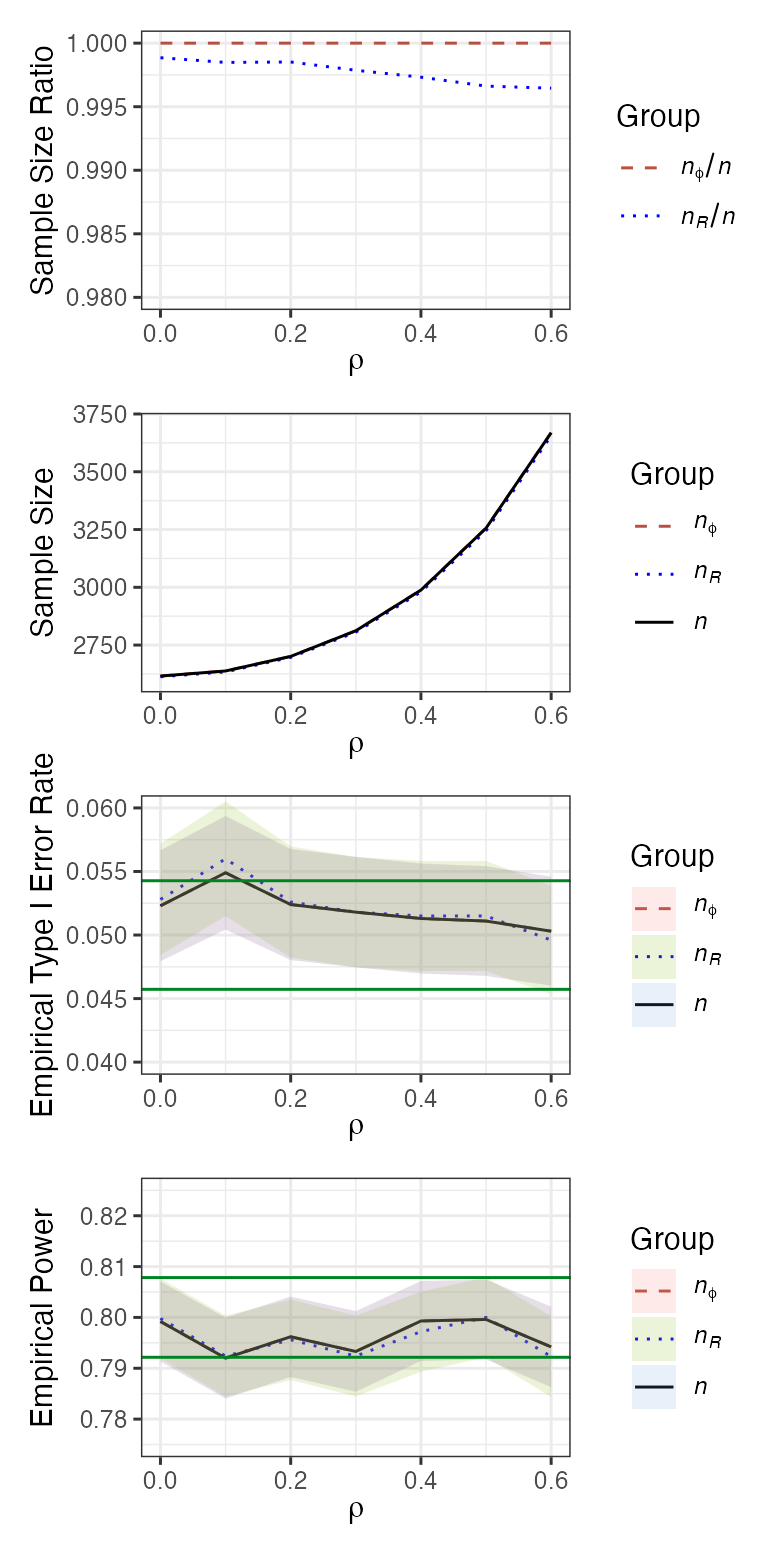}
    \caption{var $\propto$ mean$^2$}
    \end{subfigure}
    \caption{The empirical type I error rates and power of the Wald test based on the simulated count data for three cases of outcome variables, using a log link and sample sizes $n$, $n_{\phi}$, and $n_R$. The X-axis represents the correlation parameter $\rho$ used to generate the Gaussian copula. The shaded envelopes represent the 95\% confidence intervals of the empirical type I error rate and empirical power, respectively. The two horizontal lines in each panel indicate the 95\% confidence interval around the corresponding target level (0.05 for type I error rate and 0.8 for power), serving as a benchmark for good calibration. Results falling within these horizontal lines indicate well-calibrated performance.}\label{fig:count_log}
\end{figure}

Across all three cases, the 2SLiP-based sample size \( n_{\phi} \) closely matches the true  \( n \). Notably, in the third case, where \( \var(Y \mid \bz, \bx) = \sigma^2 \mu^2 \), the resulting weight \( w \) in \eqref{eq:weight} is constant, yielding \( n_{\phi} = n \). This underscores the theoretical validity of 2SLiP in settings where the weight is constant. In contrast, the P2R2 approximation \( \tRsq_{x|z} \) tends to overestimate the effect size $f^2$, and hence to underestimate the required sample size in the first two cases---``var = mean'' and ``var \( \propto \) mean''---resulting in smaller values of \( n_R \). Consequently, the achieved empirical power is lower than the nominal target, highlighting a potential limitation of P2R2 in such settings where the link function is far from linear or the variance structure is not conducive to the P2R2 approximation.

\paragraph{Identity Link}

\begin{figure}[!htb] 
    \centering
    \begin{subfigure}{.32\textwidth}
    \centering
    \includegraphics[width=\linewidth]{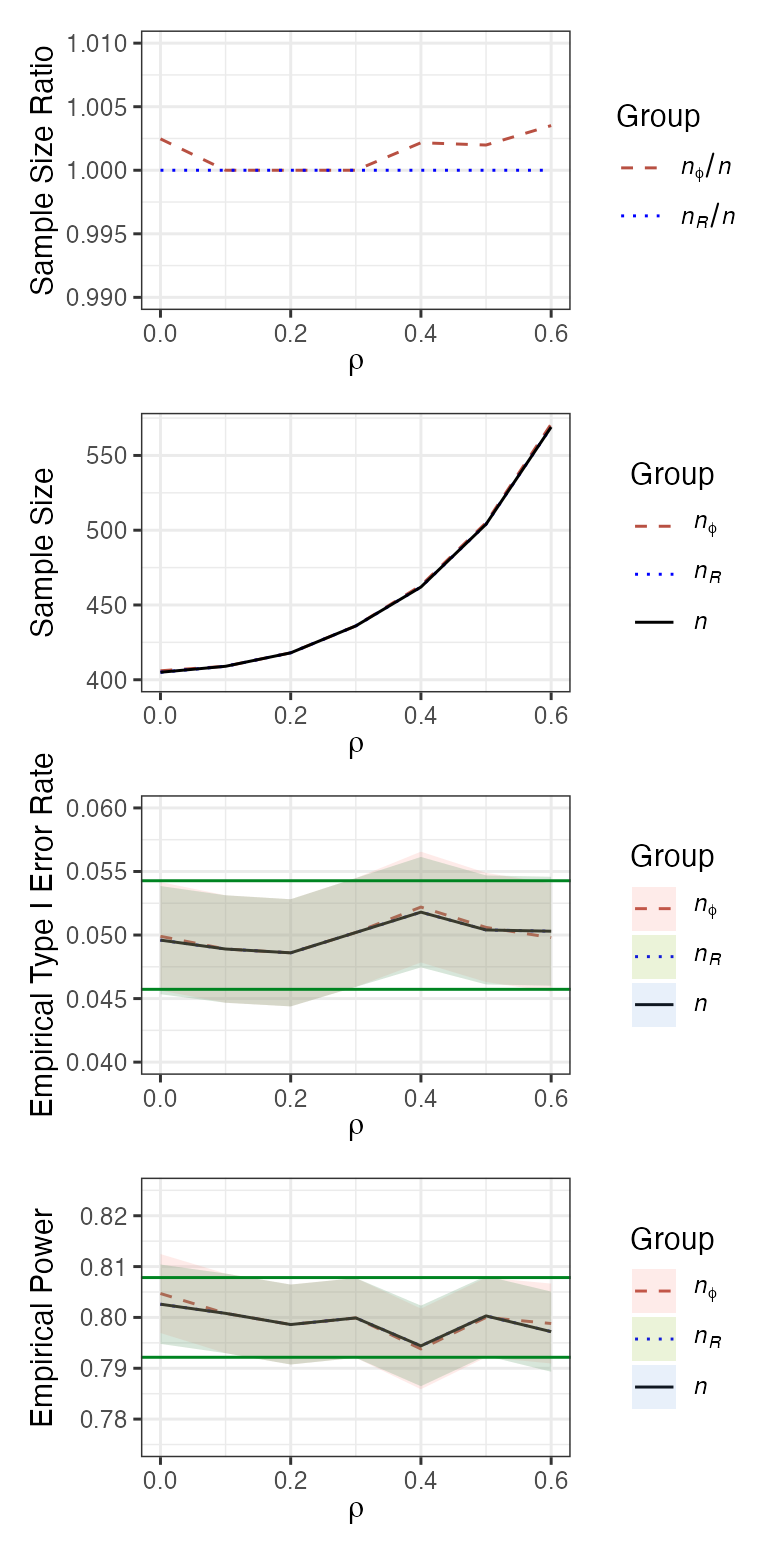}
    \caption{var = mean}
    \end{subfigure}
    \begin{subfigure}{.32\textwidth}
    \centering
    \includegraphics[width=\linewidth]{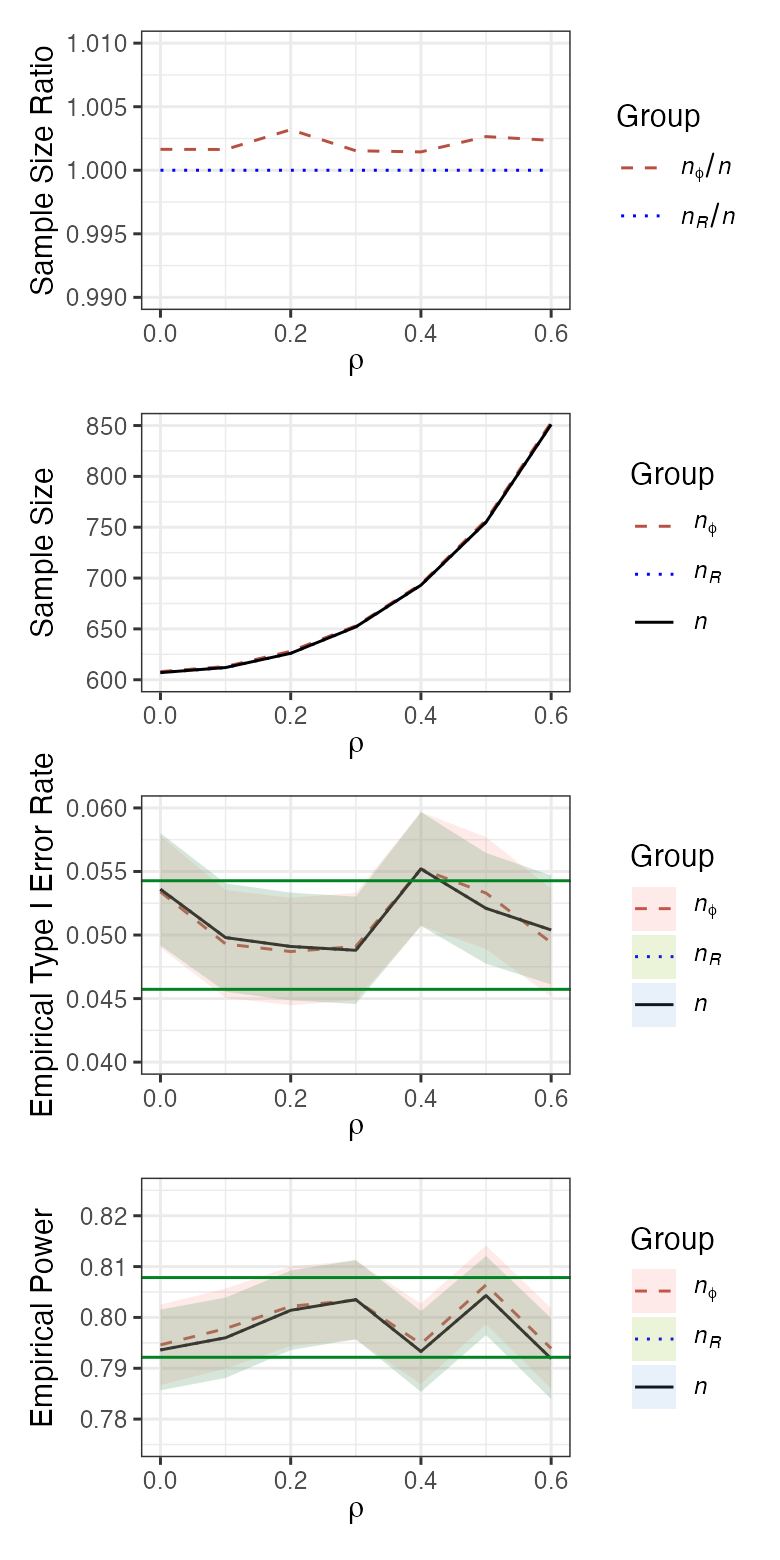}
    \caption{var $\propto$ mean}
    \end{subfigure}
    \begin{subfigure}{.32\textwidth}
    \centering
    \includegraphics[width=\linewidth]{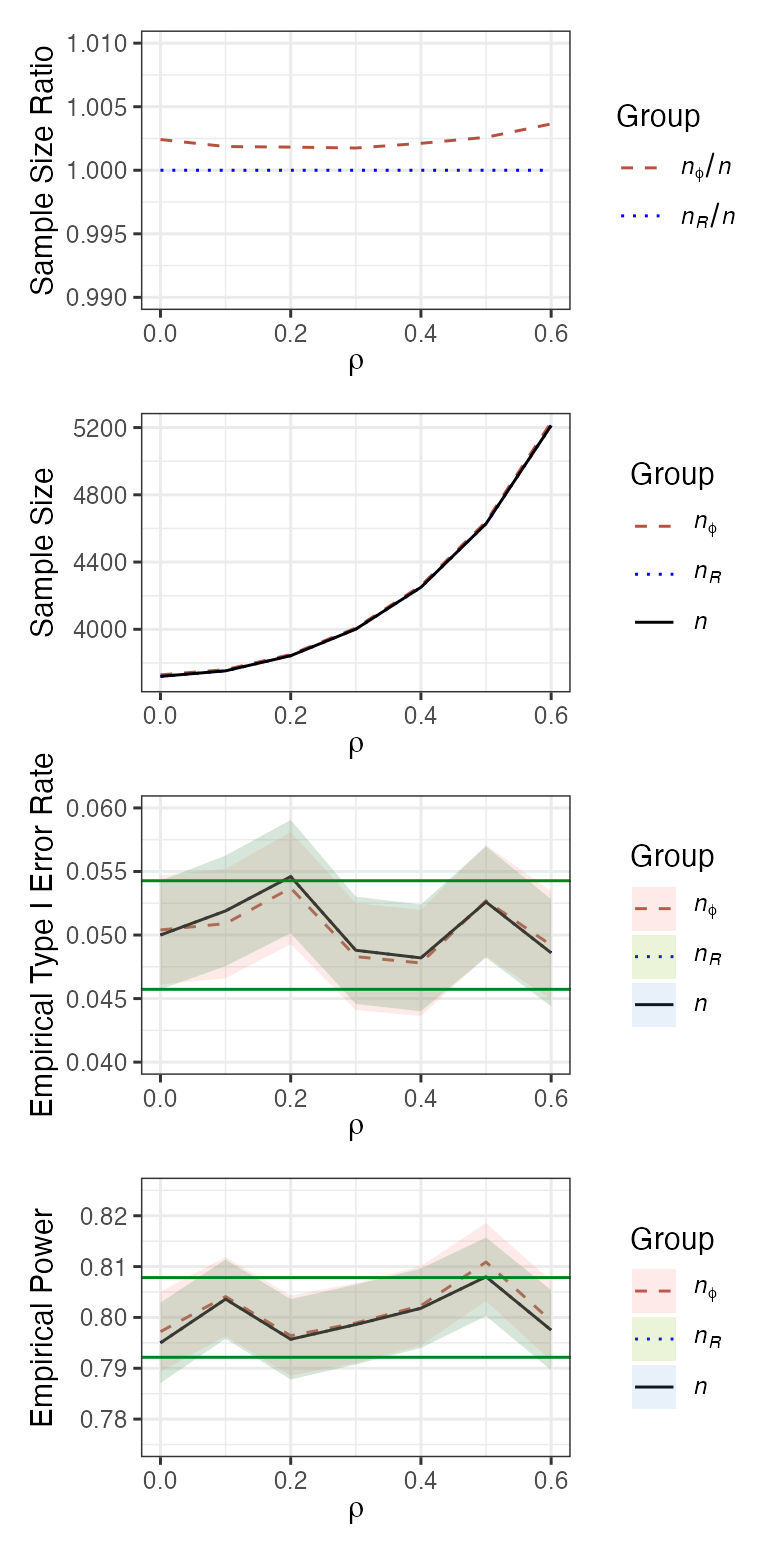}
    \caption{var $\propto$ mean$^2$}
    \end{subfigure}
    \caption{The empirical type I error rates and power of the Wald test based on the simulated count data for three cases of outcome variables, using a identity link and  sample sizes $n$, $n_{\phi}$, and $n_R$. The X-axis represents the correlation parameter $\rho$ used to generate the Gaussian copula. The shaded envelopes represent the 95\% confidence intervals of the empirical type I error rate and empirical power, respectively. The two horizontal lines in each panel indicate the 95\% confidence interval around the corresponding target level (0.05 for type I error rate and 0.8 for power), serving as a benchmark for good calibration. Results falling within these horizontal lines indicate well-calibrated performance.}\label{fig:count_identity}
\end{figure}

For the identity link, we set the coefficients $\lambda_1 = 4, \lambda_2 = 0.4, \beta_1 = 0.4$, and $\beta_2 = 0.81$, giving $\frac{d\mu}{ d \eta} = 1$.
Figure~\ref{fig:count_identity} presents the sample sizes $n$, $n_{\phi}$, and $n_{R}$ for three cases of outcome variables across  values of $\rho$, along with the empirical type I error rates and power.

Across all three cases and values of \( \rho \), the 2SLiP-based sample size \( n_{\phi} \) remains close to the true  \( n \), demonstrating robustness. Moreover, the P2R2-based sample size \( n_{R} \) equals to \( n \) exactly,  consistent with the definition of \( \tef_{R} \) and its approximation condition in \eqref{eq:mu_eta}, which requires the link function \( g(\cdot) \) to be approximately linear. Under the identity link, $\mu = \eta$ and $\mu_z = \eta_z$, so $w = 1/v$, leading to \( \tef_R =  \E[(\mu - \mu_z)^2 / v] =  \E[w \eta_{x|z}^2] = f^2\).

% \subsection{True values of $\tef$}
% Assume we know the true values of $\tef$, $\tef_{\phi}$, and $\tef_R$. To obtain them, we generate a large sample size $N$ of $Y$, $\bz$, and $\bx$, labeled as $Y_i$, $\bz_i$, and $\bx_i$, $i = 1, \cdots, N$. For example, $N = 10^6$.
% Here, $\bz_i$ and $\bx_i$ are sampled according to the process described in Section \ref{sec:rhs}. 

% For $\tef$, assuming we know $w_i$ and $\eta_{x|z,i}$, then we have
% $$\tef = \frac{1}{N}\sum_{i = 1}^N w_i \eta_{x|z,i}^2.$$
% For $\tef_{\phi}$, assuming we know $\mu_{\One} = \sum_i Y_i / N$ and $\eta_{x|z,i}$, then we know $w_{\One}$ and have
% $$\tef_{\phi} = w_{\One}\frac{1}{N}\sum_{i = 1}^N(\eta_{x|z,i} - \bar{\eta}_{x|z} )^2,$$
% where $\bar{\eta}_{x|z} = \sum_i \eta_{x|z,i} / N$.
% For $\tef_{R}$, assuming we know $v_i$, $\mu_i$ and $\mu_{z,i}$, then we have
% $$\tRsq_{x|z} = \frac{\frac{1}{N} \sum_i (\mu_i - \mu_{z,i})^2 / v_i}{1 + \frac{1}{N} \sum_i (\mu_i - \mu_{z,i})^2 / v_i}, \text{ and }\tef_{R} = \frac{\tRsq_{x|z}}{ 1 - \tRsq_{x|z}}.$$

% By plugging the expected type I error rate and power, $\alpha$ and $q$, along with the true effect $\tef$ and its alternatives $\tef_{\phi}$ and $\tef_{R}$, into equation \eqref{eq:power}, we obtain their corresponding sample sizes, $n$, $n_{\phi}$, and $n_R$. Figure \ref{fig:count_log} displays the empirical power of the Wald test based on the simulated data for three types of outcome variables, using  sample sizes $n$, $n_{\phi}$, and $n_R$. 

\subsubsection{Positive Continuous Outcomes}\label{sec:pos_cont_outcome}
We investigate two cases of outcome distributions when the outcome is a positive continuous quantity.
\textbf{First}, labeled ``var $\propto$ mean", $Y_i \sim \text{Gamma}(\mu_i/\sigma^2, \sigma^2)$, where the second parameter $\sigma^2$ is the scale. This specification yields mean  $\E[Y_i \mid \bz_i, \bx_i] = \mu_i$ and variance $\var(Y_i \mid \bz_i, \bx_i) = \sigma^2 \mu_i$. Estimation of $(\blam, \bbe)$ and $\sigma^2$ follows Section {\color{blue} S.6} in the Supplemental Materials.
\textbf{Second}, labeled ``var $\propto$ mean$^2$", assumes $Y_i \sim \text{Gamma}(1/\sigma^2, \mu_i\sigma^2)$, where the second parameter $\mu_i\sigma^2$ is the scale. This specification yields mean  $\E[Y_i \mid \bz_i, \bx_i] = \mu_i$ and variance $\var(Y_i \mid \bz_i, \bx_i) = \sigma^2 \mu_i^2$, which is a standard GLM.

\begin{figure}[!htb] 
    \centering
    \begin{subfigure}{.36\textwidth}
    \centering
    \includegraphics[width=\linewidth]{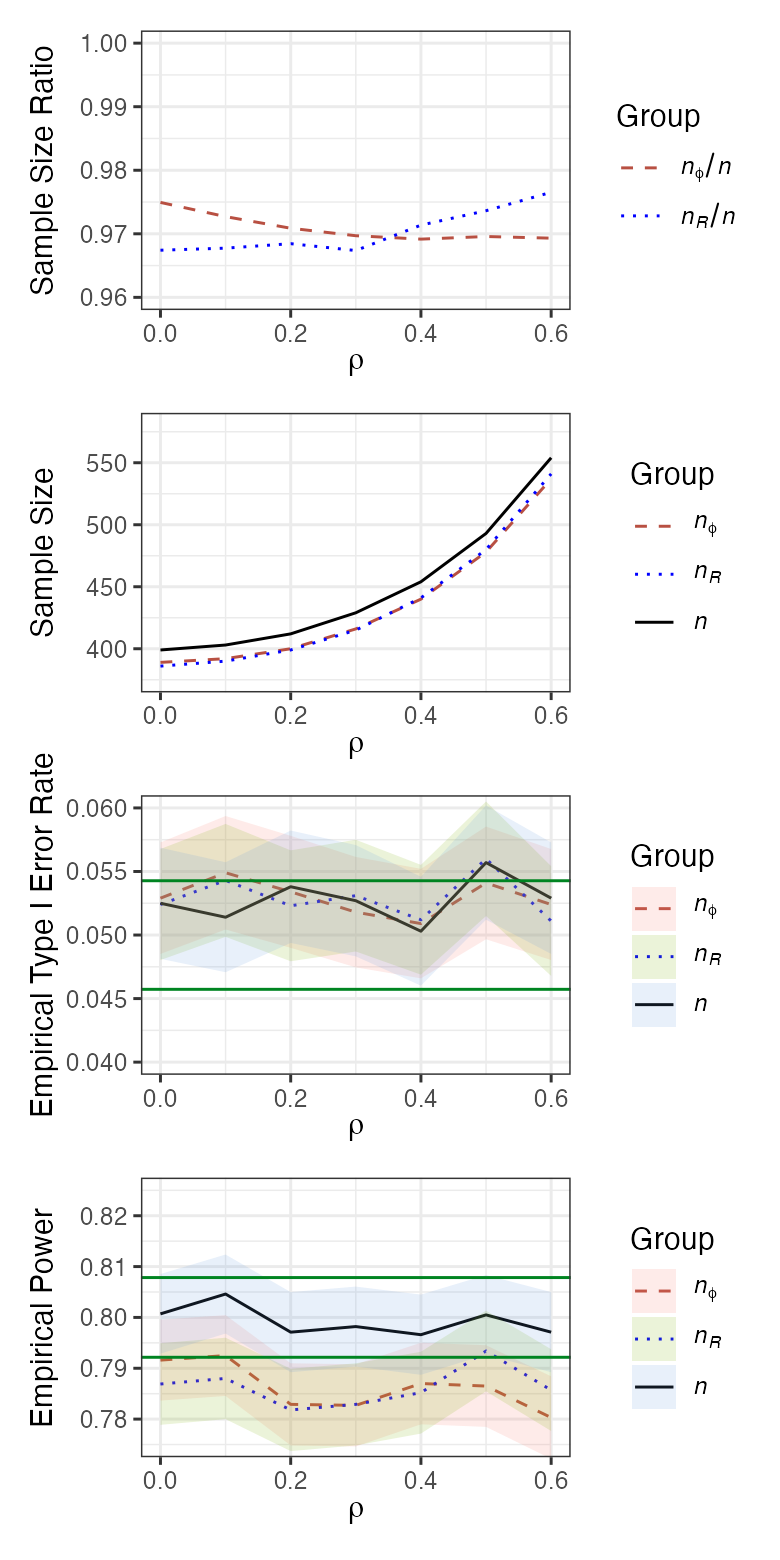}
    \caption{var $\propto$ mean} \label{fig:positive_log_1}
    \end{subfigure}
    \begin{subfigure}{.36\textwidth}
    \centering
    \includegraphics[width=\linewidth]{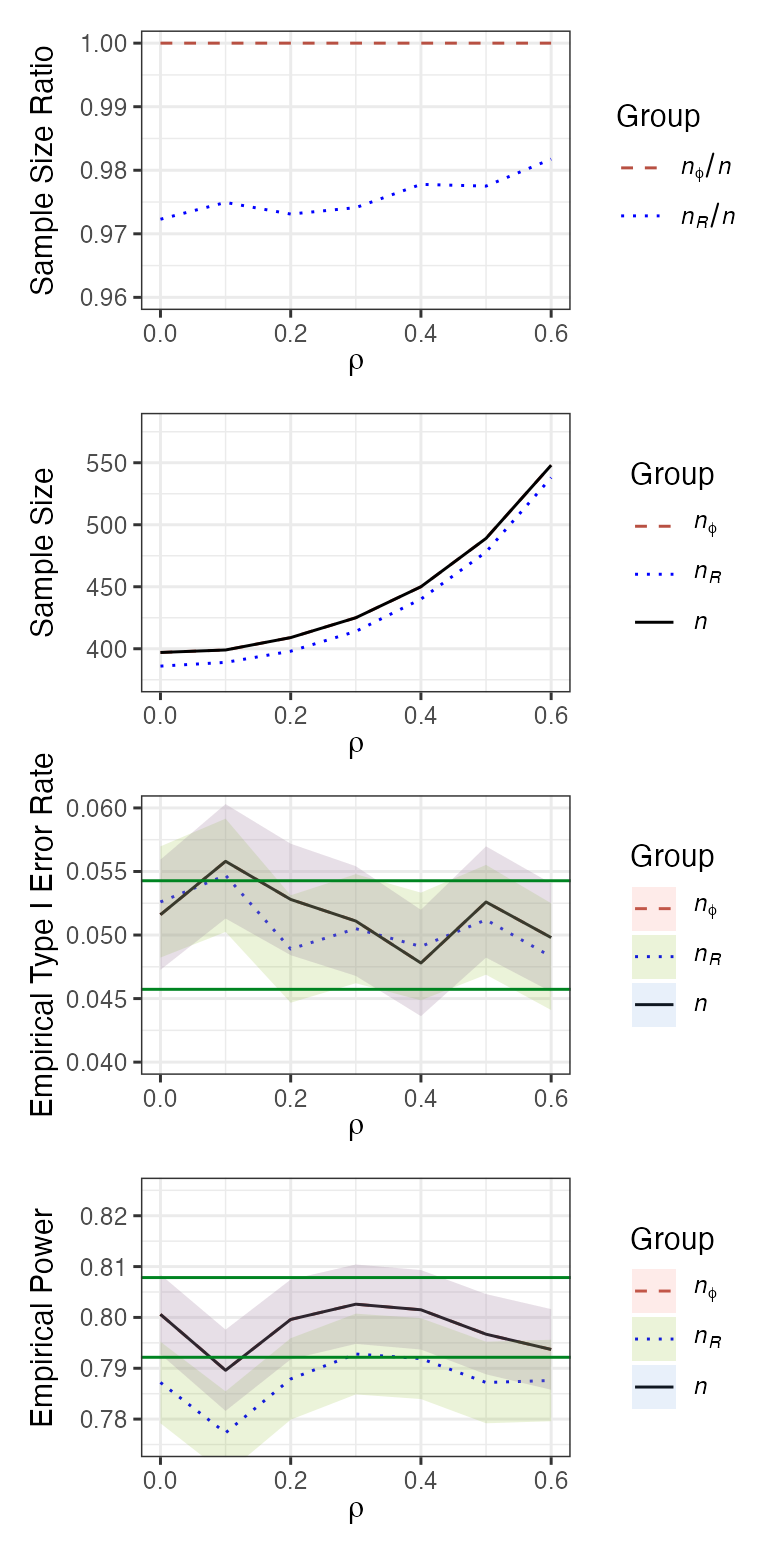}
    \caption{var $\propto$ mean$^2$}\label{fig:positive_log_2}
    \end{subfigure}
    \caption{The empirical type I error rates and power of the Wald test based on the simulated positive continuous data for two cases of outcome variables, using a log link and  sample sizes $n$, $n_{\phi}$, and $n_R$. The X-axis represents the correlation parameter $\rho$ used to generate the Gaussian copula. The shaded envelopes represent the 95\% confidence intervals of the empirical type I error rate and empirical power, respectively. The two horizontal lines in each panel indicate the 95\% confidence interval around the corresponding target level (0.05 for type I error rate and 0.8 for power), serving as a benchmark for good calibration. Results falling within these horizontal lines indicate well-calibrated performance.}\label{fig:positive_log}
\end{figure}

With targeted type I error rate $\alpha = 0.05$ and power $q = 0.8$, we set the coefficients $\lambda_1 = 1, \lambda_2 = 0.15, \beta_1 = 0.1$, and $\beta_2 = 0.15$ in this simulation study. The dispersion parameter is set to $\sigma^2 = 0.5$ for the first case and $\sigma^2 = 0.16$ for the second, both under a log link.
Figure~\ref{fig:positive_log} presents the sample sizes $n$, $n_{\phi}$, and $n_{R}$ across values of $\rho$, together with the corresponding empirical type I error rates and power.

As seen in the foregoing study, increasing $\rho$ reduces the unique contribution of $\bx$ beyond $\bz$, thereby increasing the required sample size.
In the ``var $\propto$ mean" case (Figure~\ref{fig:positive_log_1}), both $n_{\phi}$ and $n_{R}$  slightly  underestimate $n$, resulting in modest under-powering. By contrast, in the ``var $\propto$ mean$^2$" case (Figure~\ref{fig:positive_log_2}), only $n_{R}$  slightly  underestimates $n$, while \( n_{\phi} \) matches $n$ exactly. This exact equality arises for the same reason as in the third case of the log-link count outcome: when \( \var(Y \mid \bz, \bx) = \sigma^2 \mu^2 \), the resulting weight \( w \) is constant.

Across all link–variance combinations, the ratio of approximate to the true  sample size deviated from unity by no more than 3\%, with the largest under‑estimation $(n_R/n_{s} \approx 0.97)$ occurring for P2R2 in the ``var $\propto$ mean, log‑link” positive continuous case. 
Overall, these results collectively support the conclusion that 2SLiP and P2R2 offer consistently reliable performance for positive continuous outcomes. Regardless of the underlying variance structure, both measures provide valuable tools for PSS in GLMs and QL models.

 We also conduct a parallel set of simulations in which $\beta_2$ is varied while the predictor–adjustor correlation is fixed at a moderate level. These simulations provide a direct sensitivity check of how changes in effect magnitude translate into effect size and required sample size. The outcome distributions, data-generating mechanisms, estimation procedures, and Wald-test implementation are identical to those described above, and the results reinforce the robustness patterns observed in the main simulation grid (see Supplementary Section {\color{blue} S.7}. 

\subsection{Score Test Results} \label{sec:sim_res_score}

Following the setting of the simulation and in Section \ref{sec:sim_res_wald}, we target target type I error rate $\alpha = 0.05$ and power $q = 0.8$. Given $\alpha$, $q$, and degrees of freedom $p$, the true  non-centrality parameter $\Delta$ is obtained by solving equation \eqref{eq:power}. We then link $\Delta$ to the approximate score-test effect $f_s^2$ to compute the required sample size.  For a given data-generating mechanism (for example, as described at the beginning of Section \ref{sec:sim}), $f_s^2$ is evaluated  by  Monte Carlo: draw a large i.i.d. sample of  $(Y,Z,D_1,D_2)$ under known $(\blam, \bbe)$, obtain $\what{\blam}^{(0)}$ and $\what{w}^{(0)}$, and compute the sample analogue of $f_s^2$ from \eqref{eq:score_effect}. 
The required sample size is then $n_s = \lceil \Delta / f_s^2\rceil$, where $\lceil \cdot \rceil$ denotes the ceiling function. For comparison, we also report  $n_{\phi} = \lceil \Delta / \tef_{\phi} \rceil$ and $n_R = \lceil \Delta / \tef_{R} \rceil$.
Throughout this section we use $\lambda_1 = 1, \lambda_2 = 0.15, \beta_1 = 0.1$, and $\beta_2 = 0.21$ and the log link. 

Having set up and computed the required values for each case, let $N \in \{n_s, n_{\phi}, n_R\}$ be the chosen sample size. For each replicate, we generate an i.i.d.\ sample $\{(Y_i, Z_{i}, D_{1i}, D_{2i})\}_{i=1}^N$ where $Z_{i}$ and $(D_{1i}, D_{2i})$ follow Section {\color{blue} S.4} in the Supplemental Materials, and the count outcome $Y_i$ is drawn from the distributions in Section \ref{sec:count}, labled ``var = mean", ``var $\propto$ mean", and ``var $\propto$ mean$^2$". 
For each replicate and sample size, we compute the score statistic $\what{S}$
If $\what{S} > F^{-1}_{\chi^2_p(0)}(1 - \alpha)$, we reject $H_0$; otherwise, we do not. Repeating this procedure 10{,}000 times yields the empirical type I error rate under the null or the empirical power under the alternative, calculated as the proportion of simulations rejecting $H_0$.

\begin{figure}[!htb] 
    \centering
    \begin{subfigure}{.32\textwidth}
    \centering
    \includegraphics[width=\linewidth]{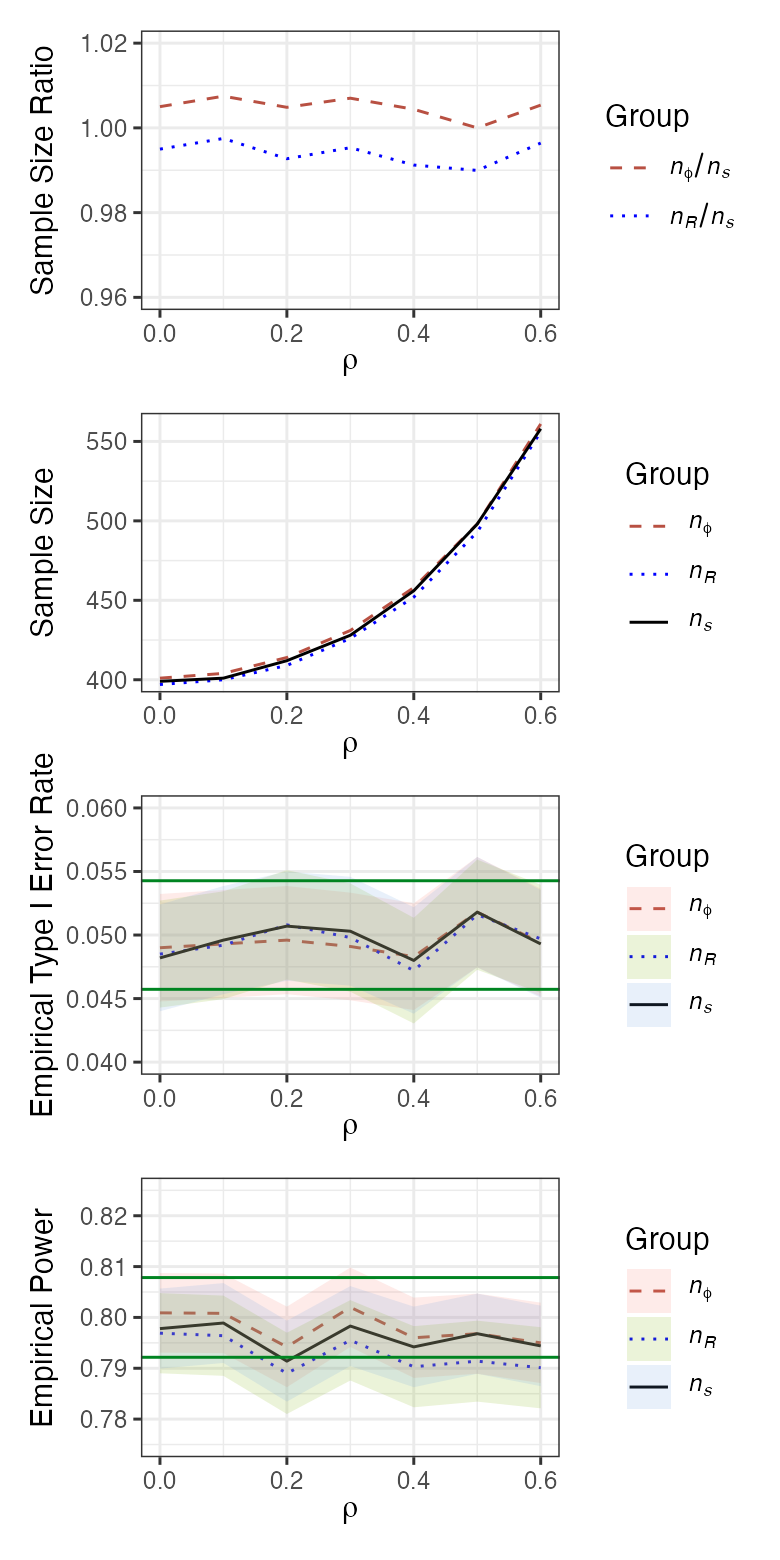}
    \caption{var = mean}
    \end{subfigure}
    \begin{subfigure}{.32\textwidth}
    \centering
    \includegraphics[width=\linewidth]{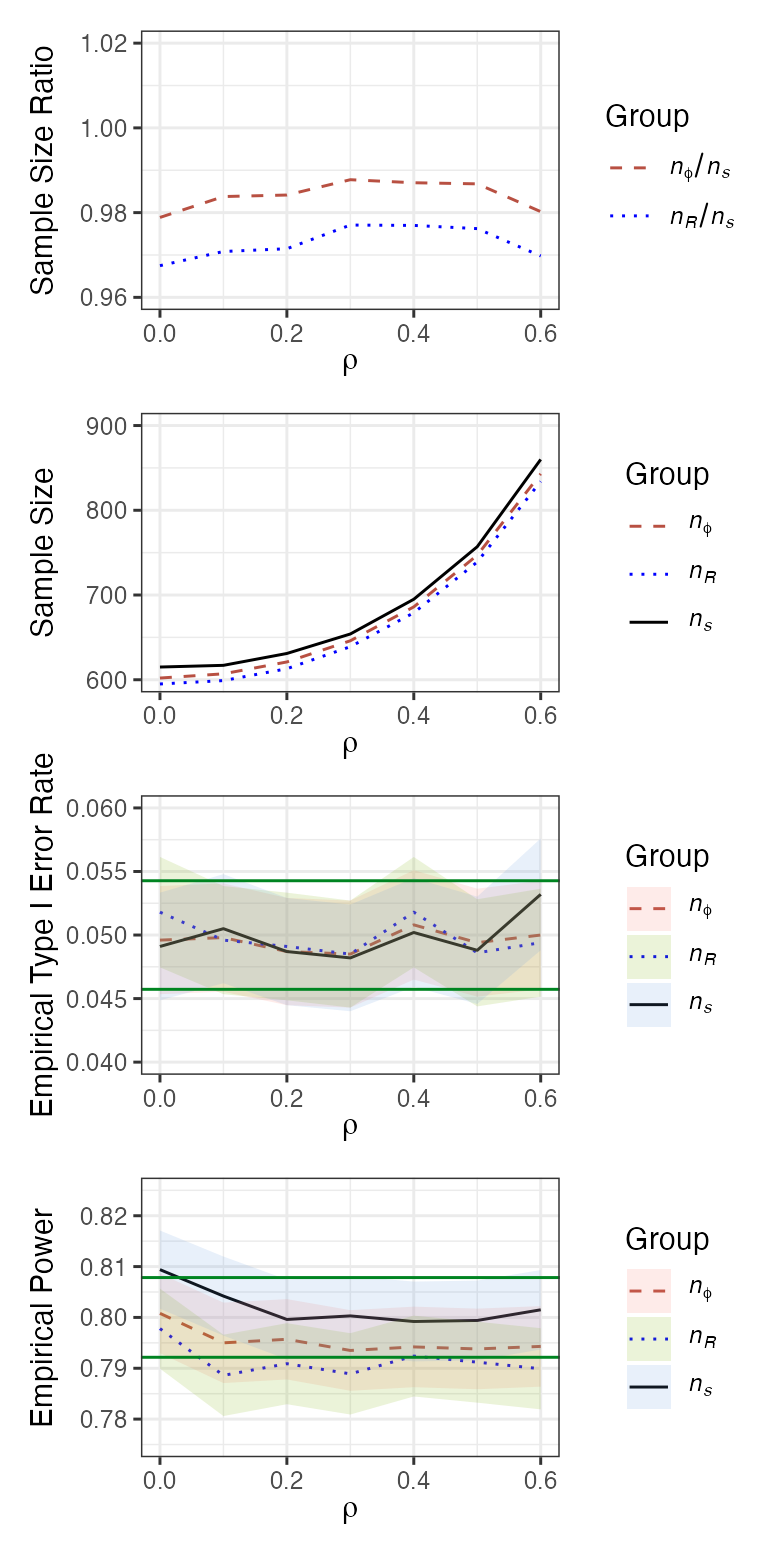}
    \caption{var $\propto$ mean}
    \end{subfigure}
    \begin{subfigure}{.32\textwidth}
    \centering
    \includegraphics[width=\linewidth]{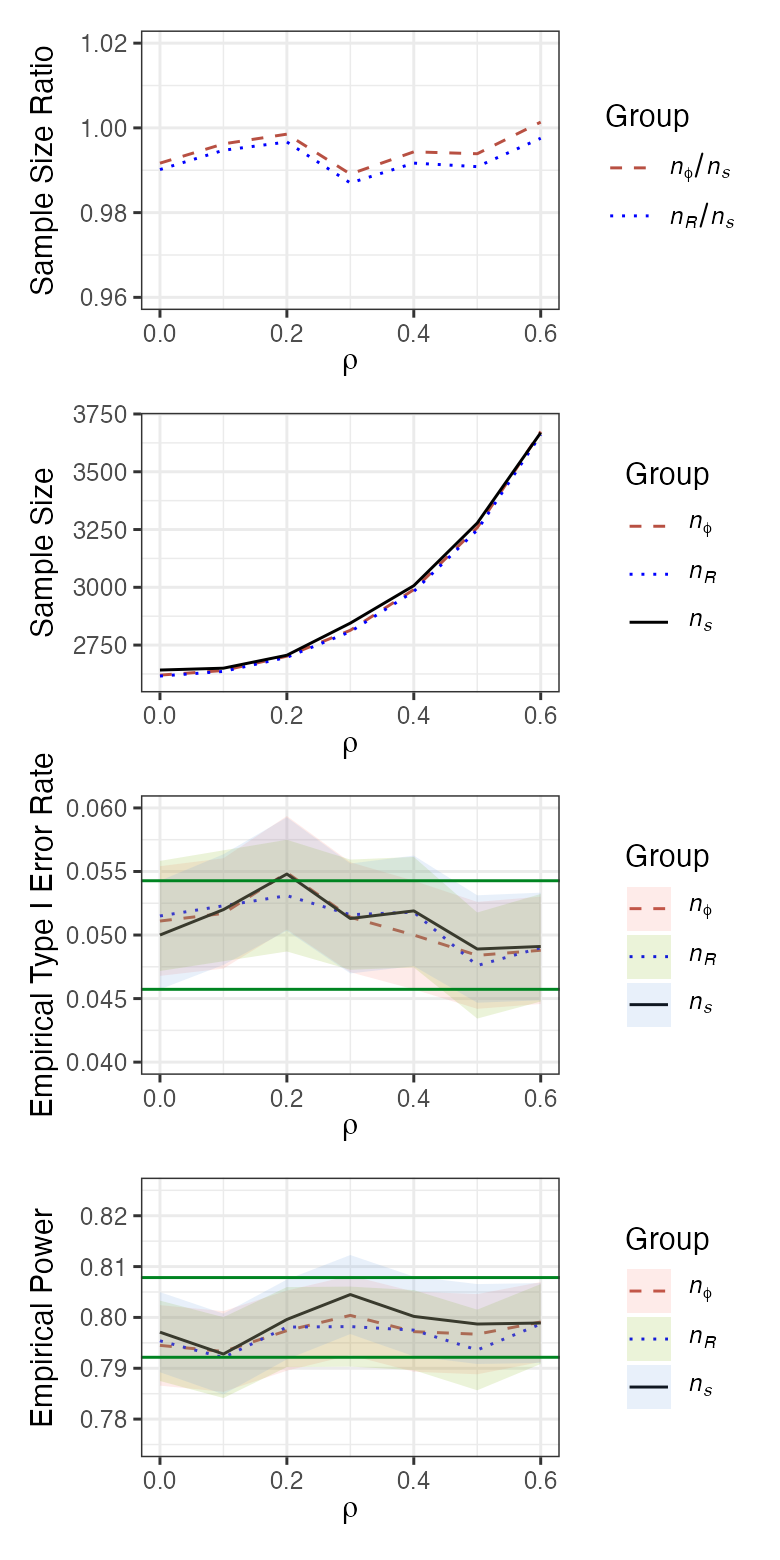}
    \caption{var $\propto$ mean$^2$}
    \end{subfigure}
    \caption{The empirical type I error rates and power of the score test based on the simulated count data for three cases of outcome variables, using a log link and sample sizes $n_s$, $n_{\phi}$, and $n_{R}$. The X-axis represents the correlation parameter $\rho$ used to generate the Gaussian copula. The shaded envelopes represent the 95\% confidence intervals of the empirical type I error rate and empirical power, respectively. The two horizontal lines in each panel indicate the 95\% confidence interval around the corresponding target level (0.05 for type I error rate and 0.8 for power), serving as a benchmark for good calibration. Results falling within these horizontal lines indicate well-calibrated performance.}\label{fig:score_count_log}
\end{figure}

Figure~\ref{fig:score_count_log} presents the sample sizes $n_s$, $n_{\phi}$, and $n_{R}$ for three outcome cases across values of the correlation parameter $\rho$, along with the empirical type I error rates and power.
Across all cases, the realized type I error remained within the target
0.05$\pm$0.004 envelope, confirming that the score statistic is well calibrated.
The sample size \(n_s\) (from the approximated score-test effect \(f_s^2\)) served as the benchmark.  
Both \(n_{\phi}\) and \(n_{R}\) matched \(n_s\) closely for ``var \(=\) mean'' and ``var \(\propto\) mean\(^2\)'', with at most 1\% deviation. 
For ``var \(\propto\) mean'', the largest underestimation was 3\%  (\(n_{\phi}/n_s = 0.98\), \(n_{R}/n_s = 0.97\)).
Increasing \(\rho\) (stronger predictor--adjustor association) inflated all three required sample sizes, but did not magnify the approximation error: the ratios \(n_{\phi}/n_s\) and \(n_{R}/n_s\) remained flat across the correlation grid.
These findings mirror those obtained for the Wald test in Section~\ref{sec:sim_res_wald}, underscoring that 2SLiP and P2R2 transfer seamlessly from Wald to score tests.

\section{Case Study} \label{sec:case}
We illustrate the two approximated effect size measures, and accompanying PSS calculations, using survey data collected from frontline health care workers during the COVID-19 pandemic \citep{cahill2022occupational}. We first analyze the data and then consider a case where these data are viewed as arising from a pilot study and show how they can be used for PSS for a future study. The dataset includes responses from participants across two health systems in central Texas and features the 23-item Burnout Assessment Tool (BAT-23) as a validated measure of burnout risk. Along with burnout outcomes, participants reported their perceptions of personal protective equipment (PPE) adequacy and provided demographic and occupational information, offering a comprehensive basis for evaluating workplace factors and mental health. In this section, we first estimated the regression coefficients, then computed the two approximated effect size measures (2SLiP and P2R2), and finally examined how variation in the hypothetical regression coefficients for the predictor influences both the effect size and the sample size requirements for future studies planned for similar designs and instruments, to be analyzed within the QL framework.

Within this case study, we applied the QL framework to model the association between perceived PPE adequacy and burnout scores. The QL approach requires only the specification of the linear predictor, link function, and variance function, making it particularly appropriate for survey-based mental health data where distributional assumptions are uncertain. In our analysis, BAT-23 scores were treated as a positive continuous outcome $Y$ and modeled with a log link $g(\cdot) = \log(\cdot)$ under a variance structure proportional to the mean. 
As model comparison is beyond the scope of this manuscript, we adopted this model framework (log link and ``var $\propto$ mean") only because it does not guarantee  exact effect size computation under either  2SLiP or P2R2, thereby allowing us to simultaneously illustrate their robustness.
Perceived PPE adequacy was entered as the predictor $\bx$, with demographic and occupational covariates (age, gender, education, and role) included as adjustors $\bz$ in the model. 
The predictor and adjustors were all categorical variables and were therefore represented using dummy variables in the regression. Specifically, perceived PPE adequacy was categorized as \textit{Very Confident}, \textit{Somewhat Confident}, \textit{Uncertain}, \textit{Somewhat Doubtful}, and \textit{Very Doubtful}. \textit{Very Confident} was treated as the reference category, and four dummy variables were created accordingly, resulting in four degrees of freedom for this predictor. 

In this model specification, the log of the conditional mean of the outcome is given by  $\log(\E[Y_i \mid \bz_i, \bx_i]) = \log(\mu_i) = \blam'\bz_i + \bbe'\bx_i$, with variance specified as $\var(Y_i \mid \bz_i, \bx_i) = \sigma^2 \mu_i$.
Using the estimation procedure described in Section {\color{blue} S.6} in the Supplemental Materials, we obtained the estimated regression coefficients for the predictor 
\begin{align*}
    \begin{array}{c c c c c c c}
        \what{\bbe} = & \big( & 0.064, &  0.100, & 0.185, & 0.142 & \big)', \\
    && \shortstack{\textit{Somewhat} \\ \textit{Confident}} & \shortstack{ \textit{Uncertain} \\ {\color{white} fill} } & \shortstack{ \textit{Somewhat} \\ \textit{Doubtful} } & \shortstack{ \textit{Very} \\ \textit{Doubtful} }
    \end{array}
\end{align*}
along with estimates for the adjustors, $\what{\blam}$, and the dispersion parameter, $\what{\sigma^2} = 0.217$.
The elements of $\what{\bbe}$ represent the log-mean differences in burnout scores relative to the \textit{Very Confident} group. 
Overall, the estimated coefficients $\what{\bbe}$ indicate that lower perceived PPE adequacy is associated with higher expected burnout scores, even after adjusting for demographic and occupational covariates.

Taking $\what{\blam}$ and $\what{\bbe}$ as the true values of $\blam$ and $\bbe$, and the empirical distribution from the data as the true distribution, we could compute $\phi_{x|z} = 0.096$, $\tRsq_{x|z} = 0.020$, the true effect size $f^2 = 0.022$ from \eqref{eq:true_effect}. 
On the original burnout scale, $\exp(\phi_{x|z}) = 1.100$ represents the expected relative change of the burnout score when perceived PPE adequacy shifts from relatively “high” confidence (e.g., \textit{Very Confident}) to relatively “low” confidence (e.g., \textit{Very Doubtful}), averaged over how workers are actually distributed across the adequacy categories and adjusting for other covariates. 
$\tRsq_{x|z}$ quantifies how much better the model predicts burnout when perceived PPE adequacy is included, beyond what’s already explained by demographic and occupational covariates. Here, $\tRsq_{x|z} = 0.020$ implies that perceived PPE adequacy explains 2.0\% of the variation in burnout scores that is not already accounted for by demographics and occupational role. Finally, $\phi_{x|z} = 0.096$ and $\tRsq_{x|z} = 0.020$ correspond to two approximated effect sizes $f^2_\phi = 0.020$ and $f^2_R = 0.021$, both of which slightly underestimate the true effect size $f^2$.

\begin{figure}[!htb] 
    \centering
    \includegraphics[width=0.65\linewidth]{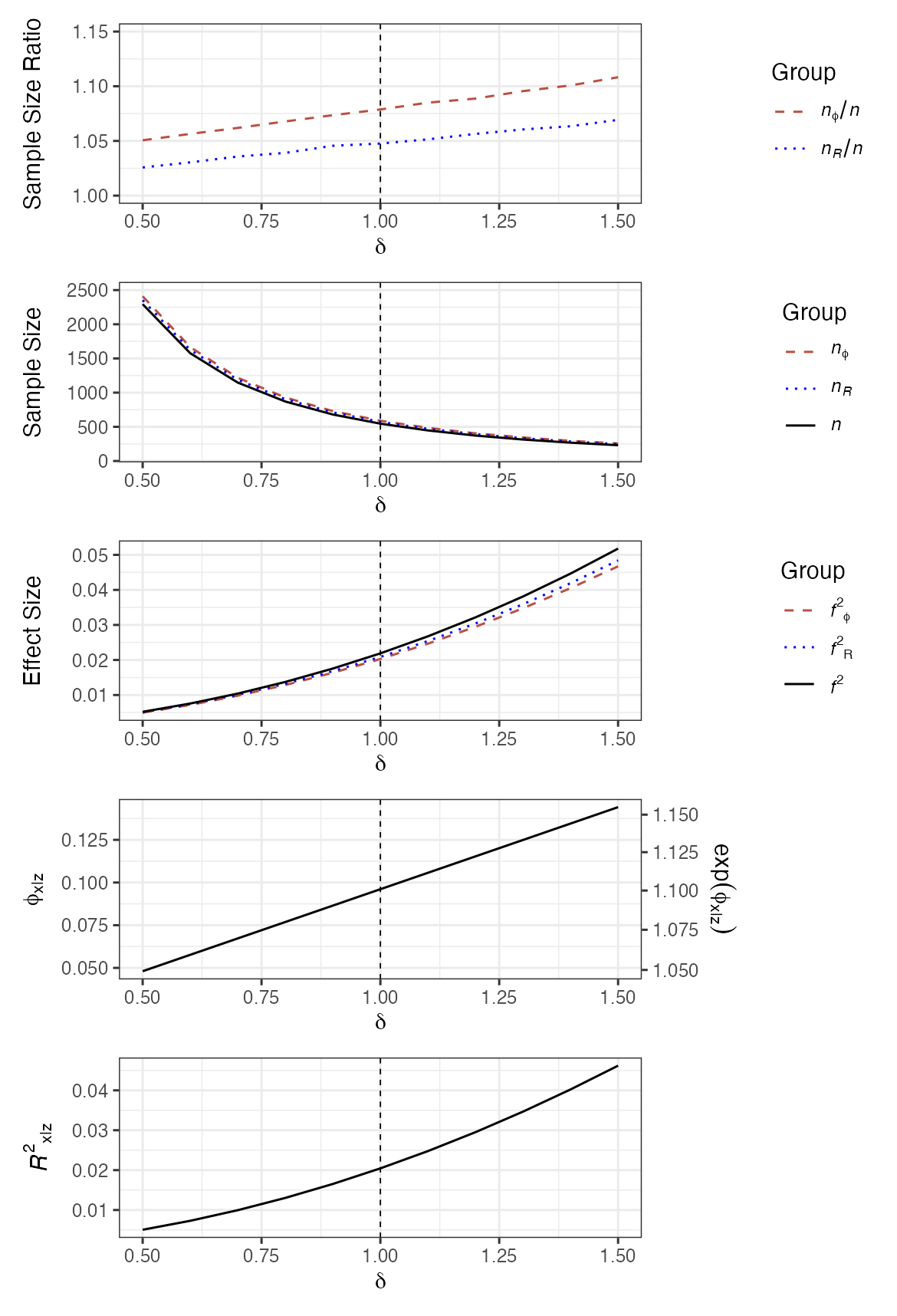}
    \caption{Performance of effect size measures and corresponding sample sizes across scaled predictor effects.}\label{fig:case}
\end{figure}

Building on this, and to show how these base values might be used as reference pilot values in study planning, we fixed $\what{\blam}$ as the true value of $\blam$ and set $\delta \cdot \what{\bbe}$ as the true value of $\bbe$, varying $\delta$ from 0.5 to 1.5 to evaluate the performance of $\phi_{x|z}$ and $\tRsq_{x|z}$. With type I error rate $\alpha = 0.05$ and degrees of freedom  $p = 4$, the true non-centrality parameter required to achieve 80\% power is $\Delta \approx 11.94$. For each value of $\delta$, we computed the true effect size $f^2$, the approximations $\phi_{x|z}$ and $\tRsq_{x|z}$ (with corresponding $f^2_\phi$ and $f^2_R$), and the approximate sample sizes required to achieve 80\% power. These sample sizes were calculated as $n = \lceil \Delta / \tef \rceil$, $n_{\phi} = \lceil \Delta / \tef_{\phi} \rceil$, and $n_R = \lceil \Delta / \tef_R \rceil$. The results are presented in Figure~\ref{fig:case}. The top panel of Figure~\ref{fig:case} shows the ratios $n_\phi/n$ and $n_R/n$, comparing sample sizes based on the approximations to those from the true effect size. Both ratios remain close to one, with deviations typically within 2--10\% and rarely exceeding 10\%. For example, when $\delta = 1$, the required sample sizes are on the order of 500--600, with $n_\phi/n = 1.079$ and $n_R/n = 1.048$, showing that both approximations remain very close to the true requirement; in this case, the approximations slightly over-estimate the needed sample size, yielding conservative study designs. Importantly, the largest deviations occur when the true required sample size is relatively small (around 200), where even moderate differences in absolute numbers translate into larger percentage discrepancies.

Overall, the case study demonstrates that both $\phi_{x|z}$ and $\tRsq_{x|z}$ yield effect size approximations that closely track the true $f^2$, providing robust and practically reliable guidance for sample size determination under the quasi-likelihood framework.

\section{Discussion} \label{sec:dis}

Using extensive simulations that span canonical and non–canonical GLMs and QL models, we found that the two effect–size measures, 2SLiP and P2R2, retain their accuracy for the Wald test, and, under more limited conditions, for the score test as well.  The 2SLiP approximation is essentially exact whenever the  weight is constant (e.g.,``var  \(\propto\) mean$^{2}$"), while P2R2 becomes exact when the link function is linear (e.g., identity link).  Even in deliberately unfavourable configurations, in our simulation investigations, both measures missed the target sample size by no more than 3\%, and the empirical power achieved by both measures deviated from the target power by no more than 2\%.

Traditional power procedures for GLMs either require full specification of the joint \((Y,\bx,\bz)\) distribution \citep{self1988power,shieh2005power,demidenko2007sample} or rely on design‑matrix enumerations that become unwieldy with many predictors \citep{lyles2007practical}.  By contrast, 2SLiP and P2R2 need only first‑ and second‑moment information, inherit the intuitive interpretation of effect in the linear predictor and variance explained from classical regression, and apply unchanged to quasi‑likelihood models with dispersion. These advantages make the new measures that we propose especially attractive when the data‑generating mechanism is incompletely specified or the design involves many predictors.

We note that soliciting 2SLiP and P2R2 inputs from investigators can still pose a challenge. Investigators must form expectations about the partial contribution of $\bx$ to the outcome, conditional on $\bz$. Additionally, both 2SLiP and P2R2 are influenced by the correlation between $\bx$ and $\bz$ and by the variance structure of $Y$, further complicating direct elicitation. However, this difficulty is no greater than that of eliciting \(R^2\) or Cohen’s \(d\) in linear regression settings, which similarly require investigators to anticipate variance explained by predictors that may not yet be fully characterized. To facilitate adoption, we are developing a user-friendly implementation of these methods as a web-based R Shiny application (\url{https://glmpower.shinyapps.io/shinyapp/}).  A follow-up manuscript describing the software resource will also include discussions and examples related to the solicitation of 2SLiP and P2R2. 

Our investigation is limited to balanced designs with categorical predictors of up to three levels. Extending the framework to highly unbalanced cells, continuous‑by‑continuous interactions, and mixed‑effects GLMs—where random effects introduce additional variance components—remains an important direction for future work. Addressing these settings will further broaden the practical scope of 2SLiP and P2R2 in planning modern, complex studies.

\section*{Declarations}

\subsection*{Ethics approval and consent to participate}
Not applicable.

\subsection*{Consent for publication}
Not applicable.

\subsection*{Availability of data and materials}
This manuscript primarily relies on simulated data to illustrate and evaluate the proposed methods. The simulated datasets and the simulation algorithms will be made publicly available upon acceptance of the manuscript. The real-world case study data are not publicly available due to confidentiality and data-use restrictions; however, the case study is included for illustrative purposes only and does not affect the main methodological conclusions of the manuscript.

\subsection*{Competing interests}
We have no conflicts of interest to
declare.

\subsection*{Funding}
The work was supported by an NIMH grant (R21MH133371).

\subsection*{Authors' contributions}
S.Y. wrote the main manuscript text and conducted the simulations. A.C. reviewed and edited the manuscript. P.R. conceived the main idea.  All authors reviewed the manuscript. We note with sadness that P.R. passed away during the final stages of this work.

\subsection*{Acknowledgements}
We used AI-based tools to assist with language polishing and editorial refinement of the manuscript. 

\bibliographystyle{plain}
\bibliography{references}

@book{mccullagh1989generalized,
  title={Generalized linear models},
  author={Mccullagh, P and Nelder, JA},
  year={1989},
  publisher={CRC press}
}

@article{hintze2006pass,
  title={{PASS user’s guide III}},
  author={Hintze, JL},
  journal={NCSS. Kaysville, Utah},
  year={2006}
}

@article{faul2007g,
  title={{G* Power 3: A flexible statistical power analysis program for the social, behavioral, and biomedical sciences}},
  author={Faul, Franz and Erdfelder, Edgar and Lang, Albert-Georg and Buchner, Axel},
  journal={Behavior research methods},
  volume={39},
  number={2},
  pages={175--191},
  year={2007},
  publisher={Springer}
}

@article{o1984procedures,
  title={Procedures for comparing samples with multiple endpoints},
  author={O'Brien, Peter C},
  journal={Biometrics},
  pages={1079--1087},
  year={1984},
  publisher={JSTOR}
}

@article{whittemore1981sample,
  title={Sample size for logistic regression with small response probability},
  author={Whittemore, Alice S},
  journal={Journal of the American Statistical Association},
  volume={76},
  number={373},
  pages={27--32},
  year={1981},
  publisher={Taylor \& Francis}
}

@article{wilson1986calculating,
  title={Calculating sample sizes in the presence of confounding variables},
  author={Wilson, SR and Gordon, I},
  journal={Journal of the Royal Statistical Society: Series C (Applied Statistics)},
  volume={35},
  number={2},
  pages={207--213},
  year={1986},
  publisher={Wiley Online Library}
}

@article{gatsonis1989multiple,
  title={Multiple correlation: exact power and sample size calculations.},
  author={Gatsonis, Constantine and Sampson, Allan R},
  journal={Psychological bulletin},
  volume={106},
  number={3},
  pages={516},
  year={1989},
  publisher={American Psychological Association}
}

@article{self1988power,
  title={Power/sample size calculations for generalized linear models},
  author={Self, Steven G and Mauritsen, Robert H},
  journal={Biometrics},
  pages={79--86},
  year={1988},
  publisher={JSTOR}
}

@article{self1992power,
  title={Power calculations for likelihood ratio tests in generalized linear models},
  author={Self, Steven G and Mauritsen, Robert H and Ohara, Jill},
  journal={Biometrics},
  pages={31--39},
  year={1992},
  publisher={JSTOR}
}

@article{signorini1991sample,
  title={Sample size for Poisson regression},
  author={Signorini, David F},
  journal={Biometrika},
  volume={78},
  number={2},
  pages={446--450},
  year={1991},
  publisher={Oxford University Press}
}

@article{hsieh1998simple,
  title={A simple method of sample size calculation for linear and logistic regression},
  author={Hsieh, Fushing Y and Bloch, Daniel A and Larsen, Michael D},
  journal={Statistics in Medicine},
  volume={17},
  number={14},
  pages={1623--1634},
  year={1998},
  publisher={Wiley Online Library}
}

@article{shieh2001sample,
  title={{Sample size calculations for logistic and Poisson regression models}},
  author={Shieh, Gwowen},
  journal={Biometrika},
  volume={88},
  number={4},
  pages={1193--1199},
  year={2001},
  publisher={Oxford University Press}
}

@article{shieh2005power,
  title={{On power and sample size calculations for Wald tests in generalized linear models}},
  author={Shieh, Gwowen},
  journal={Journal of Statistical Planning and Inference},
  volume={128},
  number={1},
  pages={43--59},
  year={2005},
  publisher={Elsevier}
}

@article{lyles2007practical,
  title={A practical approach to computing power for generalized linear models with nominal, count, or ordinal responses},
  author={Lyles, Robert H and Lin, Hung-Mo and Williamson, John M},
  journal={Statistics in Medicine},
  volume={26},
  number={7},
  pages={1632--1648},
  year={2007},
  publisher={Wiley Online Library}
}

@article{demidenko2007sample,
  title={Sample size determination for logistic regression revisited},
  author={Demidenko, Eugene},
  journal={Statistics in Medicine},
  volume={26},
  number={18},
  pages={3385--3397},
  year={2007},
  publisher={Wiley Online Library}
}

@article{demidenko2008sample,
  title={Sample size and optimal design for logistic regression with binary interaction},
  author={Demidenko, Eugene},
  journal={Statistics in Medicine},
  volume={27},
  number={1},
  pages={36--46},
  year={2008},
  publisher={Wiley Online Library}
}

@article{novikov2010modified,
  title={A modified approach to estimating sample size for simple logistic regression with one continuous covariate},
  author={Novikov, I and Fund, N and Freedman, LS},
  journal={Statistics in Medicine},
  volume={29},
  number={1},
  pages={97--107},
  year={2010},
  publisher={Wiley Online Library}
}

@article{bush2015sample,
  title={Sample size determination for logistic regression: A simulation study},
  author={Bush, Stephen},
  journal={Communications in Statistics-Simulation and Computation},
  volume={44},
  number={2},
  pages={360--373},
  year={2015},
  publisher={Taylor \& Francis}
}

@article{cahill2022occupational,
  title={{Occupational risk factors and mental health among frontline health care workers in a large US metropolitan area during the COVID-19 pandemic}},
  author={Cahill, Alison G and Olshavsky, Megan E and Newport, D Jeffrey and Benzer, Justin and Chambers, Kelly M and Custer, James and Rathouz, Paul J and Nutt, Stephanie and Jwaied, Sama and Leslie, Ryan and others},
  journal={The Primary Care Companion for CNS Disorders},
  volume={24},
  number={2},
  pages={40038},
  year={2022},
  publisher={Physicians Postgraduate Press, Inc.}
}

@article{cochran2025glm,
      title={General measures of effect size to calculate power and sample size for Wald tests with generalized linear models}, 
      author={Amy L Cochran and Shijie Yuan and Paul J Rathouz},
      year={2025},
      eprint={2506.22324},
      journal={arXiv preprint, https://arxiv.org/abs/2506.22324}
}

@article{hauck1977wald,
  title={Wald's test as applied to hypotheses in logit analysis},
  author={Hauck Jr, Walter W and Donner, Allan},
  journal={Journal of the American Statistical Association},
  volume={72},
  number={360a},
  pages={851--853},
  year={1977},
  publisher={Taylor \& Francis}
}

@article{cohen1992power,
  title={A power primer},
  author={Cohen, Jacob},
  journal={Psychological Bulletin},
  volume={112},
  number={1},
  pages={155--159},
  year={1992}
}

@article{channouf2014power,
  title={Power and sample size calculations for Poisson and zero-inflated Poisson regression models},
  author={Channouf, Nabil and Fredette, Marc and MacGibbon, Brenda},
  journal={Computational Statistics \& Data Analysis},
  volume={72},
  pages={241--251},
  year={2014},
  publisher={Elsevier}
}

@article{wedderburn1974quasi,
  title={Quasi-likelihood functions, generalized linear models, and the Gauss—Newton method},
  author={Wedderburn, Robert WM},
  journal={Biometrika},
  volume={61},
  number={3},
  pages={439--447},
  year={1974},
  publisher={Oxford University Press}
}

@article{mccullagh1983quasi,
  title={Quasi-likelihood functions},
  author={McCullagh, Peter},
  journal={The Annals of Statistics},
  pages={59--67},
  year={1983},
  publisher={JSTOR}
}

@article{igeta2018power,
  title={Power and sample size calculation incorporating misspecifications of the variance function in comparative clinical trials with over-dispersed count data},
  author={Igeta, Masataka and Takahashi, Kunihiko and Matsui, Shigeyuki},
  journal={Biometrics},
  volume={74},
  number={4},
  pages={1459--1467},
  year={2018},
  publisher={Wiley Online Library}
}

\begin{appendices}
\appendixpage
\setcounter{table}{0}
\renewcommand{\thetable}{A.\arabic{table}}
\setcounter{equation}{0}
\renewcommand{\theequation}{A.\arabic{equation}}
\setcounter{figure}{0}
\renewcommand{\thefigure}{A.\arabic{figure}}
\renewcommand{\thealgorithm}{A.\arabic{algorithm}}

\section{Asymptotic Justification of the Wald Statistic under Quasi-Likelihood} \label{sec:supp-wald-asymptotic}
We begin by approximating the expected information matrix $ \bi_{n}$ with the marginal information matrix $\mI$. As $\what{\blam}$  and $\what{\bbe}$ are consistent estimators of $\blam^*$ and $\bbe^*$, and assuming that $(\bx_i,\bz_i)$ are i.i.d., the Law of Large Number gives $\bi_n\left(\what{\blam}, \what{\bbe}\right)/ n   \xrightarrow{p} \mI$, where the weight $w$ in $\mI$ is evaluated at the true QL parameter values of $\blam^*$ and $\bbe^*$.
Hence, by \cite{mccullagh1983quasi}, 
$$
\sqrt{n}\left(\begin{bmatrix}
\what{\blam} \\
\what{\bbe}
\end{bmatrix} - 
\begin{bmatrix}
\blam^* \\
\bbe^* 
\end{bmatrix}\right)
\overset{A}{\sim} \mathcal{N}\left( 
\bo,
\mI^{-1}
% \left\{\bi_n\left(\what{\blam}, \what{\bbe}\right)\right\}^{-1}
\right).
$$
For $\what{\bbe}$, $\bi_{n, \beta|\lambda}\left(\what{\blam}, \what{\bbe}\right)/ n   \xrightarrow{p} \cIbl$ and $\sqrt{n} \left(\what{\bbe} -  \bbe^* \right) \overset{A}{\sim} \mathcal{N}\left(  \bo,\cIbl^{-1} \right)$, 
% Similarly, we have $\bi_{n,\bbe|\blam}^{-1}/n \xrightarrow{p} \cIbl^{-1}$, 
where $\cIbl^{-1}$ is the lower right $p \times p$ block of $\mI^{-1}$ given by the Schur complement:
$$\cIbl \equiv \E\left[ w \bx \bx' \right] - \E\left[ w \bx \bz' \right] \E\left[ w \bz \bz' \right]^{-1} \E\left[ w \bz \bx' \right]. $$

For testing the null hypothesis $H_0: \bbe^* = 0$, the Wald test statistic is given by
\[
\what{W} =  \what{\bbe}' \bi_{n,\bbe|\blam}\left(\what{\blam}, \what{\bbe}\right) \what{\bbe},
\]
which can be approximated as
\[
\what{W} \approx n \what{\bbe}' \cIbl  \what{\bbe}.
\]
Under the asymptotic normality of $\sqrt{n}(\what{\bbe} - \bbe^*) \xrightarrow{d} N(\bo, \cIbl^{-1})$, we write:
$$\what{\bbe} = \bbe^* + \frac{1}{\sqrt{n}}\boldsymbol{O}, \text{ where } \boldsymbol{O} \sim N(\bo, \cIbl^{-1}).$$
Substituting into $\what{W}$, we obtain:
$$
\begin{aligned}
    \what{W} \approx & n \left(\bbe^* + \frac{1}{\sqrt{n}}\boldsymbol{O}\right)' \cIbl  \left(\bbe^* + \frac{1}{\sqrt{n}}\boldsymbol{O}\right)\\
    = & n \left[{\bbe^*}'\cIbl\bbe^* + \frac{2}{\sqrt{n}} {\bbe^*}'\cIbl\boldsymbol{O} + \frac{1}{n}\boldsymbol{O}'\cIbl\boldsymbol{O}\right] \\
    = & n {\bbe^*}'\cIbl\bbe^* + 2\sqrt{n} {\bbe^*}'\cIbl\boldsymbol{O} + \boldsymbol{O}'\cIbl\boldsymbol{O}. \\
\end{aligned}
$$

Take the expectation of $\what{W}$. The first term is deterministic; the second term  has expectation zero since $\E[\boldsymbol{O}] = \bo$; the third term evaluates to:
$$
\E\left[\boldsymbol{O}'\cIbl\boldsymbol{O}\right] = \text{tr}\left(\cIbl \E\left[\boldsymbol{O}\boldsymbol{O}'\right]\right) = \text{tr}\left(\cIbl \cIbl^{-1}\right) = p.
$$
Thus, the expectation of the Wald statistic is $\E\left[\what{W}\right] =  n {\bbe^*}'\cIbl\bbe^* + p$. This implies that the non-centrality parameter is $\Delta = n {\bbe^*}'\cIbl\bbe^* = n f^2$, and the corresponding effect
$$f^2 = {\bbe^*}'\cIbl\bbe^*.$$

\section{Theoretical Justification of Effect Size–Based Power Approximations} \label{sec:supp-power-calculations}
The preceding derivation in Section \ref{sec:supp-wald-asymptotic} establishes the asymptotic behavior of the Wald statistic from an information-based perspective; we next offer a complementary explanation by exploiting the representation of QLEs as solutions to iteratively weighted least squares.
Specifically, The QLEs \(\what{\blam}\) and \(\what{\bbe}\) can be obtained from \(n\) independent observations of \((Y, \bx, \bz)\) by iteratively constructing a linearized working variable 
$$g(Y) \approx Y_l \equiv \eta + \frac{d\eta}{d \mu} (Y - \mu),$$
and QLEs $(\what{\blam},\what{\bbe})$ result from regressing \(Y_l\) onto \((\bx,\bz)\) with the weight \(w\), updating until convergence. Equivalently, this can be written as
\begin{equation} \label{eq:weighted_lm}
    w^{1/2} Y_l = w^{1/2} \eta + \frac{(Y - \mu)}{\sqrt{v}} = \blam' w^{1/2}\bz + \bbe' w^{1/2}\bx + \frac{(Y - \mu)}{\sqrt{\sigma^2 v}}.
\end{equation}

To further disentangle the contributions of $\bx$ and $\bz$, 
% the Frisch–Waugh–Lovell theorem \citep{frisch1933partial,lovell1963seasonal} can be applied. Because $\bx$ may contain  components that are collinear with $\bz$, 
we regress $w^{1/2}\bx$ on $w^{1/2}\bz$ to extract the residual component of $\bx$ that is orthogonal to $\bz$ in the weighted space, leaving the residual $w^{1/2}(\bx - \bA  \bz)$, where $\bA_{p \times r}$ denotes the population limit of the weighted least squares estimation of the regression coefficients from regressing $\bx$ on $\bz$, defined as
\begin{equation} \label{eq:A}
    \bA = \E[w\bx\bz']E[w\bz\bz']^{-1}.
\end{equation}
Substituting this decomposition into \eqref{eq:weighted_lm} yields
% This allows $\eta$ to be decomposed into a part explained by  $\bz$ and a residual part of $\bx$ beyond $\bz$.
% Let $\bA_{p \times r}$ denote the population limit of the weighted least squares estimation of the regression coefficients, defined as
% where $w$ is computed using the true parameter values of $\blam$ and $\bbe$.
% Thus, \eqref{eq:weighted_lm} can be rewritten as
\begin{equation} \label{eq:eta_decomp}
w^{1/2} Y_l = w^{1/2}(\blam' + \bbe' \bA)\bz + w^{1/2} (\bbe'\bx - \bbe'\bA \bz) + \frac{(Y - \mu)}{\sqrt{\sigma^2 v}}.
\end{equation}

As our focus is on interpretation rather than estimation, we specialize to the true QL parameter values $\blam^*$ and $\bbe^*$. In this case, the linear predictor admits the decomposition $\eta = \eta_z + \eta_{x|z}$, where 
\begin{equation} \label{eq:reduced_linear_predictor}
    \eta_z = ({\blam^*}' + {\bbe^*}' \bA)\bz
\end{equation}
captures the contribution of the adjustors $\bz$, and 
$$\eta_{x|z} = {\bbe^*}' (\bx - \bA\bz)$$
represents the unique contribution of the residualized $\bx$, scaled by~$w^{1/2}$. Consequently,
$$
w^{1/2} Y_l = w^{1/2} \eta_z + w^{1/2}\eta_{x|z} + \frac{(Y - \mu)}{\sqrt{\sigma^2 v}}.
$$

It follows that the true effect $f^2$ can be equivalently expressed as
$$
\tef = {\bbe^*}' \E[w(\bx - \bA\bz)(\bx - \bA\bz)'] \bbe^* = \E\left[w\eta_{x|z}^2\right],
$$
since
\[
\begin{aligned}
    \tef = & {\bbe^*}' \E[w(\bx - \bA\bz)(\bx - \bA\bz)'] \bbe^* \\
    = & {\bbe^*}' \E[w(\bx\bx' - \bx\bz'\bA' - \bA\bz\bx' + \bA\bz\bz'\bA)] \bbe^* \\
    = & {\bbe^*}' \left( \E[w\bx\bx'] - \E[w\bx\bz']\bA'  - \bA\E[w\bz\bx'] + \bA\E[w\bz\bz']\bA' \right) \bbe^* \\
    = & {\bbe^*}' \left( \E[w\bx\bx'] - \E[w\bx\bz']E[w\bz\bz']^{-1}\E[w\bz\bx']\right) \bbe^* \\
    = & {\bbe^*}' \cIbl \bbe^*. \\
\end{aligned}
\]
The fourth equality holds because 
$$\E[w\bx\bz']\bA' =  \bA\E[w\bz\bx'] = \bA\E[w\bz\bz']\bA' = \E[w\bx\bz']E[w\bz\bz']^{-1}\E[w\bz\bx'].$$
Throughout, the expectation, variance, and covariance operators are taken over the joint distribution of $\bx$ and $\bz$, with the weight 
$w$ evaluated at the the true QL parameter values $\blam^*$ and $\bbe^*$.

\paragraph{Approximation of True Effect}
Expansion of  $\E\left[w\eta_{x|z}^2\right]$ yields
$$
\tef = \E[w]\E\left[\eta_{x|z}^2\right] + \cov\left[w, \eta_{x|z}^2\right] = \E[w] \left(E[\eta_{x|z}]^2 + \var[\eta_{x|z}]\right) + \cov\left[w, \eta_{x|z}^2\right].
$$
The first measure $\phi_{x|z}$ captures a 2SD change in the residualized linear predictor $\eta_{x|z}$. Under local conditions where $\E\left[\eta_{x|z}\right] \approx 0$, $\cov\left[w,\eta_x^2\right] \approx 0$, and $\E[w]\approx w_{\One}$,
the true effect $\tef$ can be approximated by $ \tef_{\phi} \equiv  w_{\One} \phi_{x|z}^2/4 \approx \tef,$
where $w_{\One}$ denotes the value of $w$ when evaluated at the population mean response $\mu = \E[Y]$, marginally w.r.t $\bx$ and $\bz$.

The connection between $\tef$ and $\tRsq_{x|z}$ becomes clear through a first-order approximation. For small differences between \( \eta \) and \( \eta_z \),
$$
\mu - \mu_z \approx \frac{d\mu}{d\eta} (\eta - \eta_z),
$$
which implies $\tef_R \equiv \tRsq_{x|z} / (1 - \tRsq_{x|z}) \approx \E[w \eta_{x|z}^2] = \tef.$
Thus $\tRsq_{x|z}$ connects directly to the non-centrality parameter in Wald tests, particularly when the link is nearly linear.

\section{Equivalence of Reduced-Model Linear Predictors}

The reduced-model linear predictor $\eta_z$ is introduced from two perspectives:
(i) as the solution to a weighted mean squared error (WMSE) minimization problem in \eqref{eq:min_wmse_reduced} involving only the adjustors $\bz$, and
(ii) as the component of the full linear predictor  \eqref{eq:reduced_linear_predictor} obtained by partialling out $\bx$ with respect to $\bz$ in the weighted space.
Here we show that these two constructions are equivalent.

Recall that the reduced-model coefficients are defined as
\[
\blam_0^*
=
\argmin_{\blam}
\E\!\left[
w\{Y_l-\blam' \bz\}^2
\right],
\]
where the working variable $Y_l$ and weight $w$ are evaluated at the quasi-true parameter values $(\blam^*,\bbe^*)$ of the full model.
The first-order condition for this minimization problem is
\[
\E[w\bz\bz']\blam
=
\E[w\bz Y_l].
\]

Under the working linearization at the true QL parameter values, the linearized outcome admits the decomposition
\[
Y_l = ({\blam^*}' + {\bbe^*}' \bA)\bz + ({\bbe^*}'\bx - {\bbe^*}'\bA \bz) + \frac{d\eta}{d \mu} (Y - \mu).
\]
To establish the equivalence, we substitute the above decomposition of $Y_l$ into the first-order condition $\E[w\bz Y_l].$
By linearity of expectation, this yields three corresponding terms, which we examine in turn.

\paragraph{First term.}
Consider
\[
\E\!\left[
w \bz \big\{ ({\blam^*}' + {\bbe^*}' \bA)\bz \big\}
\right]
=
\E[w\bz\bz'](\blam^*+\bA'\bbe^*).
\]

\paragraph{Second term.}
Next, consider
\[
\E\!\left[
w \bz \big\{ {\bbe^*}'\bx - {\bbe^*}'\bA \bz \big\}
\right]
=
\E[w\bz\bx']\bbe^* - \E[w\bz\bz']\bA'\bbe^*.
\]
By the definition of $\bA$ in \eqref{eq:A}, 
it follows that
\[
\E[w\bz\bx'] = \E[w\bz\bz']\bA',
\]
and hence this term is identically zero.

\paragraph{Third term.}
Finally, consider the contribution from the working error term,
\[
\E\!\left[
w \bz \frac{d\eta}{d \mu} (Y - \mu)
\right].
\]
Under the QL assumption $\E[Y\mid \bx,\bz]=\mu$, we have
\[
\E\!\left[ \frac{d\eta}{d \mu} (Y - \mu) \mid \bx,\bz \right]=0,
\]
and therefore, by iterated expectation,
\[
\E\!\left[
w \bz \frac{d\eta}{d \mu} (Y - \mu)
\right]=\bo.
\]

Combining the three components, we obtain
\[
\E[w\bz Y_l]
=
\E[w\bz\bz'](\blam^*+\bA'\bbe^*).
\]
Substituting this expression into the first-order condition yields
\[
\E[w\bz\bz']\blam
=
\E[w\bz\bz'](\blam^*+\bA'\bbe^*),
\]
and, assuming $\E[w\bz\bz']$ is nonsingular,
\[
\blam_0^* = \blam^*+\bA'\bbe^*.
\]
Consequently, the reduced-model linear predictor satisfies
\[
\eta_z = {\blam_0^*}'\bz = ({\blam^*}'+{\bbe^*}'\bA)\bz,
\]
which coincides with the representation obtained by partialling out $\bx$ with respect to $\bz$ in the weighted space. This establishes the equivalence of the two constructions of $\eta_z$.

\section{RHS Data Generation} \label{sec:rhs_DG}
To introduce a dependency between the continuous uniform variable \(Z\) and a categorical variable \(X\) with three levels \(\{0, 1, 2\}\), we use a Gaussian copula.
The steps for generating \(Z\) and \(X\) using a Gaussian copula are as follows:

\begin{enumerate}
    \item Generate correlated normal variables \((C_1, C_2)\) from the bivariate normal distribution with mean zero and covariance matrix \(\Sigma\):
    \[
    (C_1, C_2) \sim N\left( 
    \begin{bmatrix}
    0 \\ 
    0
    \end{bmatrix}, 
    \Sigma \right),
    \]
    where
    \[
    \Sigma = 
    \begin{bmatrix}
    1 & \rho \\
    \rho & 1
    \end{bmatrix},
    \]
    and \(\rho \in [-1, 1]\) is the correlation parameter between $C_1$ and $C_2$. 
    \item Transform $C_1$ and $C_2$ to two uniform variables via the cumulative distribution function (CDF) of the standard normal distribution $\Phi(\cdot)$:
    \[
    U_1 = \Phi(C_1), \quad U_2 = \Phi(C_2).
    \]
    \item Set \(Z = U_1\).
    \item Define the categorical variable \(X\) by the interval into which \(U_2\) falls:
    \[
    X = 
    \begin{cases}
    0 & \text{if} \quad U_2 \in [0, 1/3), \\
    1 & \text{if} \quad U_2 \in [1/3, 2/3), \\
    2 & \text{if} \quad U_2 \in [2/3, 1].
    \end{cases}
    \]
\end{enumerate}

\section{Coefficient Search}\label{sec:coeff_search}

\begin{algorithm}
\caption{Adjust \(\beta_2\) to match target sample size \(n^\star\)}
\begin{algorithmic}[1]
\REQUIRE Target non-centrality parameter \(\Delta\), target sample size \(n^\star\), joint distribution of \((Z, D_1,D_2)\), coefficients \(\lambda_1, \lambda_2, \beta_1\), initial \(\beta_2 \in (-\infty, \infty)\), tolerance \(\epsilon\), step size \(\kappa\) 
\ENSURE Estimated \(\beta_2\) such that \(n \approx n^\star\)

\WHILE{not converged}
    \STATE Compute effect size \(\tef = \tef(\lambda_1, \lambda_2, \beta_1, \beta_2)\) using Monte Carlo methods
    \STATE Compute \(n = \frac{\Delta}{\tef}\)
    \IF{\(|n - n^\star| < \epsilon\)}
        \STATE \textbf{break}
    \ELSIF{\(n > n^\star\)}
        \STATE \(\beta_2 := \beta_2 - \kappa\)
    \ELSE
        \STATE \(\beta_2 := \beta_2 + \kappa\)
    \ENDIF
\ENDWHILE

\RETURN \(\beta_2\)
\end{algorithmic}
\end{algorithm}

\clearpage\newpage
\section{Quasi-Likelihood Estimation Algorithm}\label{sec:quasi_est}
For simplicity, we use \( \bx \) to represent all covariates $\bx$ and $\bz$ in the main manuscript, and let \( \bbe \) denote their corresponding coefficients in the estimation algorithm. The parameter $\sigma^2$ denotes the dispersion. 

In quasi-likelihood theory \citep{mccullagh1989generalized}, if conditional mean and variance of $Y$ given $\bx$ are $\E[Y \mid \bx] = \mu$ and $\var(Y \mid \bx) = v$, then the quasi-likelihood function $Q(Y;\bbe,\sigma^2)$ satisfies
$$\frac{\partial Q(Y;\bbe,\sigma^2)}{\partial\mu} =  \frac{Y - \mu}{v}$$
%  = \frac{Y - \mu}{\sigma^2 \mu^2}.
The quasi-score function is the derivative of the quasi-likelihood with respect to the parameter vector $\bbe$,
$$\bU(\bbe,\sigma^2) = \frac{\partial Q(Y;\bbe,\sigma^2)}{\partial\bbe} = \frac{\partial Q(Y;\bbe,\sigma^2)}{\partial\mu} \frac{d\mu}{d\eta} \frac{\partial\eta}{\partial\bbe} = \frac{Y - \mu}{v} \frac{d\mu}{d\eta} \bx$$

Given \( N \) observations \( \{Y_i, \bx_i\} \), we iteratively estimate the coefficients  using weighted least squares to solve the quasi-score equation
\begin{equation}\label{eq:score_eq0}
    \bU_N(\bbe,\sigma^2) = \sum_{i=1}^N \frac{Y_i - \mu_i}{v_i} \left(\frac{d\mu}{d\eta} \Big|_{\eta = \eta_i}\right) \bx_i = 0.
\end{equation}
During the iterative procedure, we update the dispersion parameter \( \sigma^2 \) using the Pearson estimate. 

Algorithm \ref{algorithm_quasi} applies to the third case of count outcome in Section \ref{sec:count}, which follows a modified negative binomial distribution. In this setting, the dispersion parameter $\sigma^2$ and the variance $v_i$ are updated as follows:
\[
{\sigma^2}^{(t+1)} = \frac{1}{N - p} \sum_{i=1}^N \frac{(Y_i - \mu_i^{(t)})^2}{\left( \mu_i^{(t)} \right)^2}
\quad \text{ and } \quad 
v_i^{(t)} = {\sigma^2}^{(t)} (\mu_i^{(t)})^2
\]
This algorithm can be readily adapted to estimate the coefficients and dispersion parameter for the first case of positive continuous outcome in Section \ref{sec:pos_cont_outcome} by modifying the updates as follows:
\[
{\sigma^2}^{(t+1)} = \frac{1}{N - p} \sum_{i=1}^N \frac{(Y_i - \mu_i^{(t)})^2}{\mu_i^{(t)}}
\quad \text{ and } \quad 
v_i^{(t)} = {\sigma^2}^{(t)} \mu_i^{(t)}
\]

% \[ \left( v_i^{(t)} = {\sigma^2}^{(t)} (\mu_i^{(t)})^2 \text{ for negative binomial and } v_i^{(t)} = {\sigma^2}^{(t)} \mu_i^{(t)} \text{ for gamma} \right) \]

% \( \left( {\sigma^2}^{(t+1)} = \frac{1}{N - p} \sum_{i=1}^N \frac{(Y_i - \mu_i^{(t)})^2}{\left( \mu_i^{(t)} \right)^2} \text{ for negative binomial and } {\sigma^2}^{(t+1)} = \frac{1}{N - p} \sum_{i=1}^N \frac{(Y_i - \mu_i^{(t)})^2}{ \mu_i^{(t)} } \text{ for gamma} \right) \)

\begin{algorithm}
\footnotesize
\caption{Iterative Quasi-Likelihood Estimation} \label{algorithm_quasi}
\begin{algorithmic}[1]
\REQUIRE \( N \) observations \( \{Y_i, \bx_i\} \) (design matrix $\bx_{(N)} = \begin{pmatrix}\bx_1 & \cdots & \bx_N\end{pmatrix}'$), link function $g(\cdot)$, initial regression coefficients \( \bbe^{(0)} \) and dispersion parameter \( {\sigma^2}^{(0)} \), tolerance \(\epsilon\)
\ENSURE Estimated \(\bbe\) and \(\sigma^2\) satisfying the quasi-score equation \eqref{eq:score_eq0}
\FOR{iteration \( t = 0, 1, 2, \ldots \)}
    \FOR{each observation \( i = 1, \ldots, N \)}
        \STATE Compute linear predictor: \( \eta_i^{(t)} = \bbe^{(t)'} \bx_i \)
        \STATE Compute mean: \( \mu_i^{(t)} = g^{-1}(\eta_i^{(t)}) \)
        % \STATE Compute variance: $v_i^{(t)}$ 
        % \( \left( v_i^{(t)} = {\sigma^2}^{(t)} (\mu_i^{(t)})^2 \text{ for negative binomial and } v_i^{(t)} = {\sigma^2}^{(t)} \mu_i^{(t)} \text{ for gamma} \right) \)
        \STATE Compute variance: \( v_i^{(t)} = {\sigma^2}^{(t)} (\mu_i^{(t)})^2 \)
        \STATE Compute working variable: 
        \[
        z_i^{(t)} = \eta_i^{(t)} + (Y_i - \mu_i^{(t)}) \cdot \left(\frac{d\eta}{d\mu} \Big|_{\mu = \mu_i^{(t)}}\right)
        \]
        \STATE Compute weight: 
        \[
        w_i^{(t)} = \frac{1}{v_i^{(t)}}  \left(\frac{d\mu}{d\eta} \Big|_{\eta = \eta_i^{(t)}}\right)^2
        \]
    \ENDFOR
    \STATE Form weight matrix \( \boldsymbol{W}^{(t)} = \mathrm{diag}(w_1^{(t)}, \ldots, w_N^{(t)}) \)
    \STATE Form working response vector \( \boldsymbol{z}^{(t)} = (z_1^{(t)}, \ldots, z_N^{(t)})' \)
    \STATE Update coefficients:
    \[
    \bbe^{(t+1)} = \left( \bx_{(N)}' \boldsymbol{W}^{(t)} \bx_{(N)} \right)^{-1} \bx_{(N)}' \boldsymbol{W}^{(t)} \boldsymbol{z}^{(t)}
    \]
    \vspace{-1em}
    \STATE Update dispersion:
    \[
    {\sigma^2}^{(t+1)} = \frac{1}{N - p} \sum_{i=1}^N \frac{(Y_i - \mu_i^{(t)})^2}{\left( \mu_i^{(t)} \right)^2}
    \]
    \IF{\( \|\bbe^{(t+1)} - \bbe^{(t)}\|_{\infty} < \epsilon \) \AND \( |{\sigma^2}^{(t+1)} - {\sigma^2}^{(t)}| < \epsilon \)}
        \STATE \textbf{break}
    \ENDIF
\ENDFOR
\end{algorithmic}
\end{algorithm}

\clearpage\newpage
\section{Additional Simulation Results}\label{sec:add_sim}

To further examine the performance of 2SLiP and P2R2 on power and sample size estimation, we conducted an additional set of simulations, fixing the correlation parameter at $\rho = 0.3$ and varying the coefficient $\beta_2$ while holding other parameters constant. This setup isolates the impact of one predictor’s effect size under a moderate degree of predictor–adjustor dependence. For each value of $\beta_2$, we computed the true non-centrality parameter $f^2$, its approximations $f^2_\phi$ and $f^2_R$, and the corresponding required sample sizes $n$, $n_\phi$, and $n_R$. The outcome distributions, data-generating mechanisms, estimation procedures, and Wald-test conduction mirror those described in Section~\ref{sec:sim_res_wald}.

\subsection{Count Outcomes} 

\subsubsection{Log Link} 
\begin{figure}[!htb] 
    \centering
    \begin{subfigure}{.32\textwidth}
    \centering
    \includegraphics[width=\linewidth]{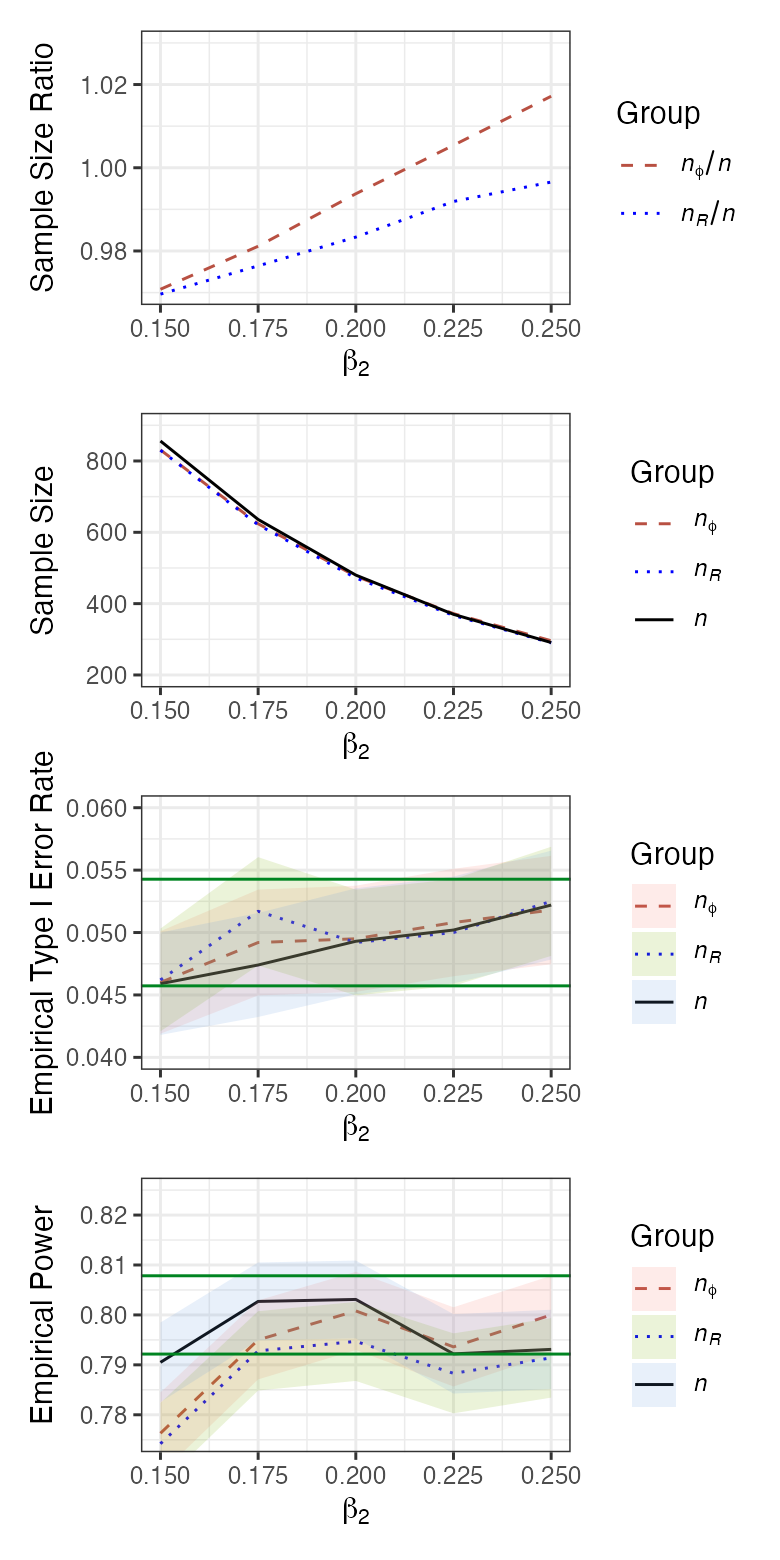}
    \caption{var = mean}
    \end{subfigure}
    \begin{subfigure}{.32\textwidth}
    \centering
    \includegraphics[width=\linewidth]{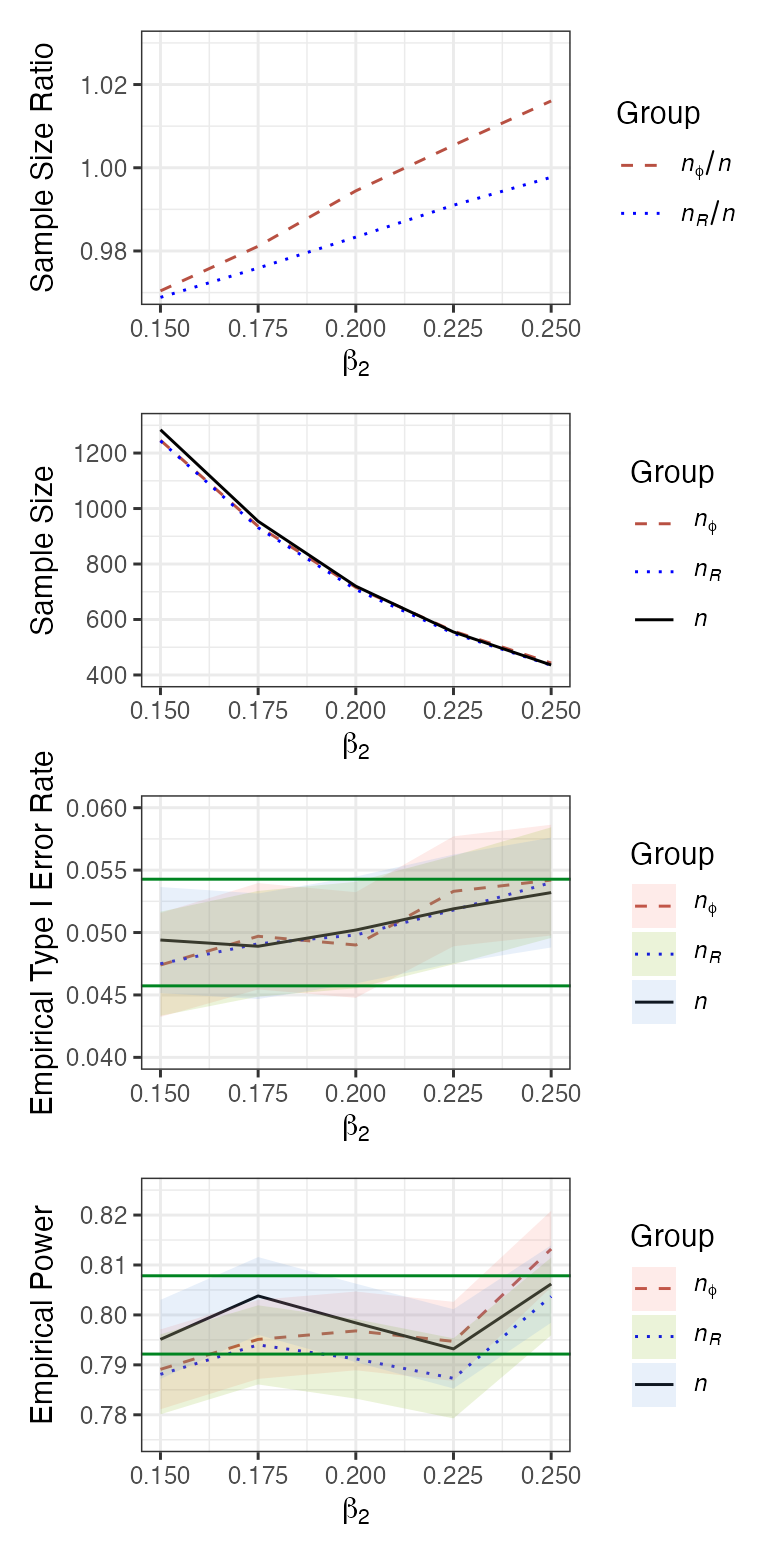}
    \caption{var $\propto$ mean}
    \end{subfigure}
    \begin{subfigure}{.32\textwidth}
    \centering
    \includegraphics[width=\linewidth]{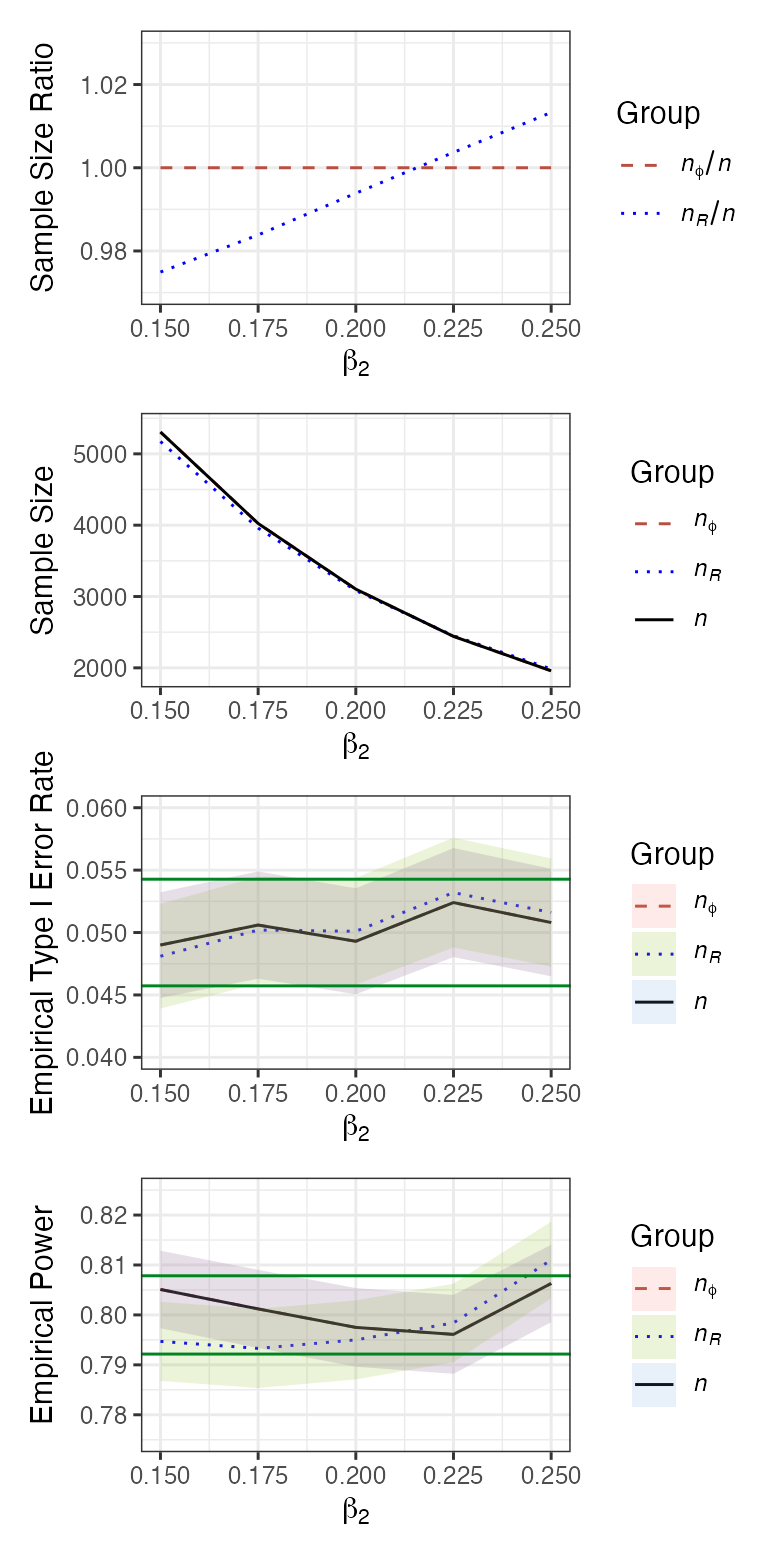}
    \caption{var $\propto$ mean$^2$}
    \end{subfigure}
    \caption{The empirical type I error rates and power of the Wald test based on the simulated count data for three cases of outcome variables, using a log link and sample sizes $n$, $n_{\phi}$, and $n_R$. The X-axis represents the coefficient $\beta_2$. The shaded envelopes represent the 95\% confidence intervals of the empirical type I error rate and empirical power, respectively. The two horizontal lines in each panel indicate the 95\% confidence interval around the corresponding target level (0.05 for type I error and 0.8 for power), serving as a benchmark for good calibration. Results falling within these horizontal lines indicate well-calibrated performance.}\label{fig:count_log_vary_beta}
\end{figure}

We fixed $\lambda_1 = 1$, $\lambda_2 = 0.15$, and $\beta_1 = 0.1$, and varied $\beta_2$ from 0.15 to 0.25 across a grid of values. For each value of $\beta_2$, we considered the same three count outcome distributions as in the main simulation study: Poisson (var = mean), over-dispersed Poisson (var $\propto$ mean), and modified negative binomial (var $\propto$ mean$^2$), each fitted using a log link. We then assessed the performance of 2SLiP and P2R2 by comparing the approximate sample sizes $n_\phi$ and $n_R$, as well as the resulting empirical power, against the true sample size $n$ and its empirical power across varying values of $\beta_2$.
Figure~\ref{fig:count_log_vary_beta} displays the simulation results under the log link.

\subsubsection{Identity Link} 
\begin{figure}[!htb] 
    \centering
    \begin{subfigure}{.32\textwidth}
    \centering
    \includegraphics[width=\linewidth]{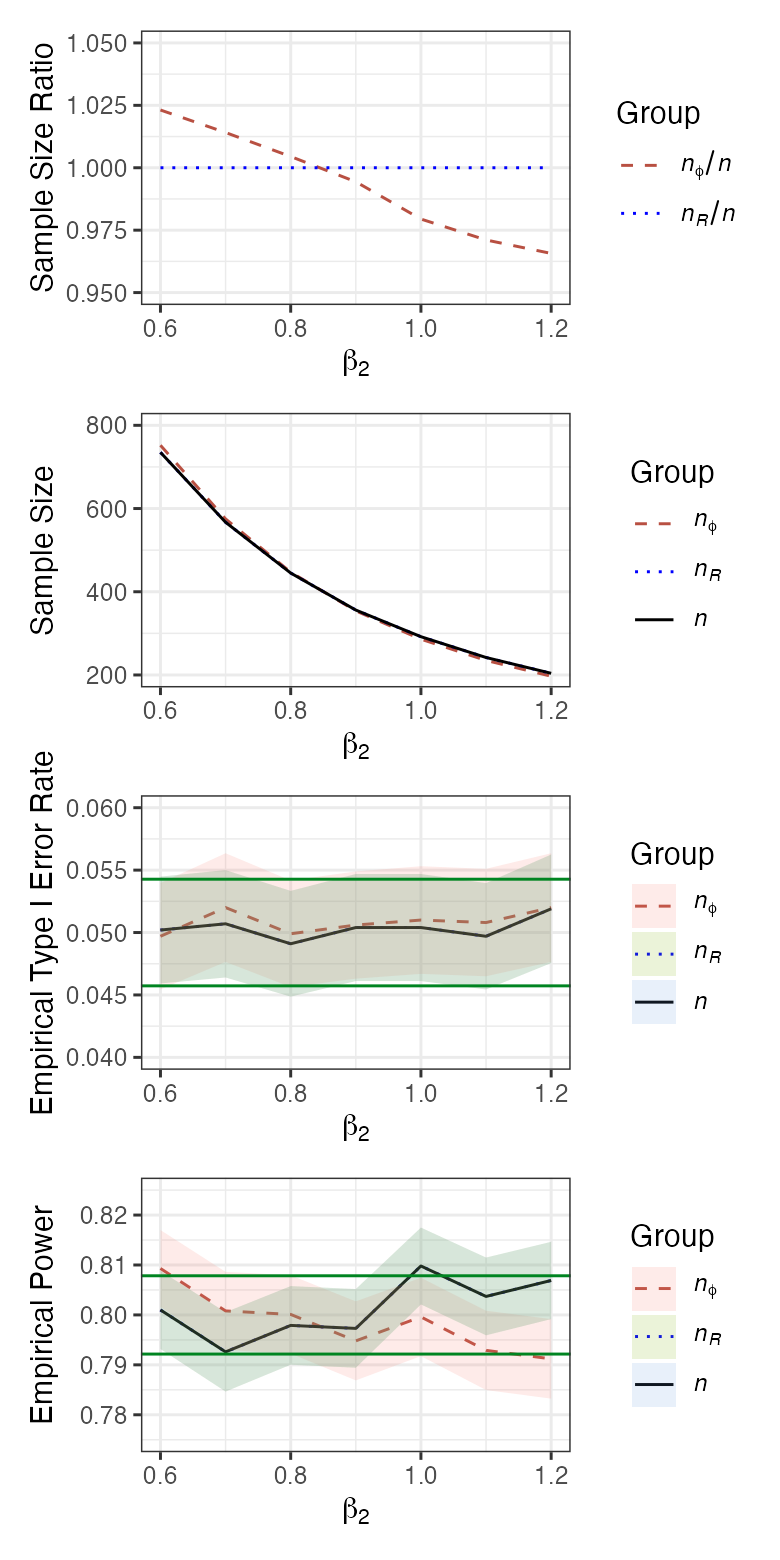}
    \caption{var = mean}
    \end{subfigure}
    \begin{subfigure}{.32\textwidth}
    \centering
    \includegraphics[width=\linewidth]{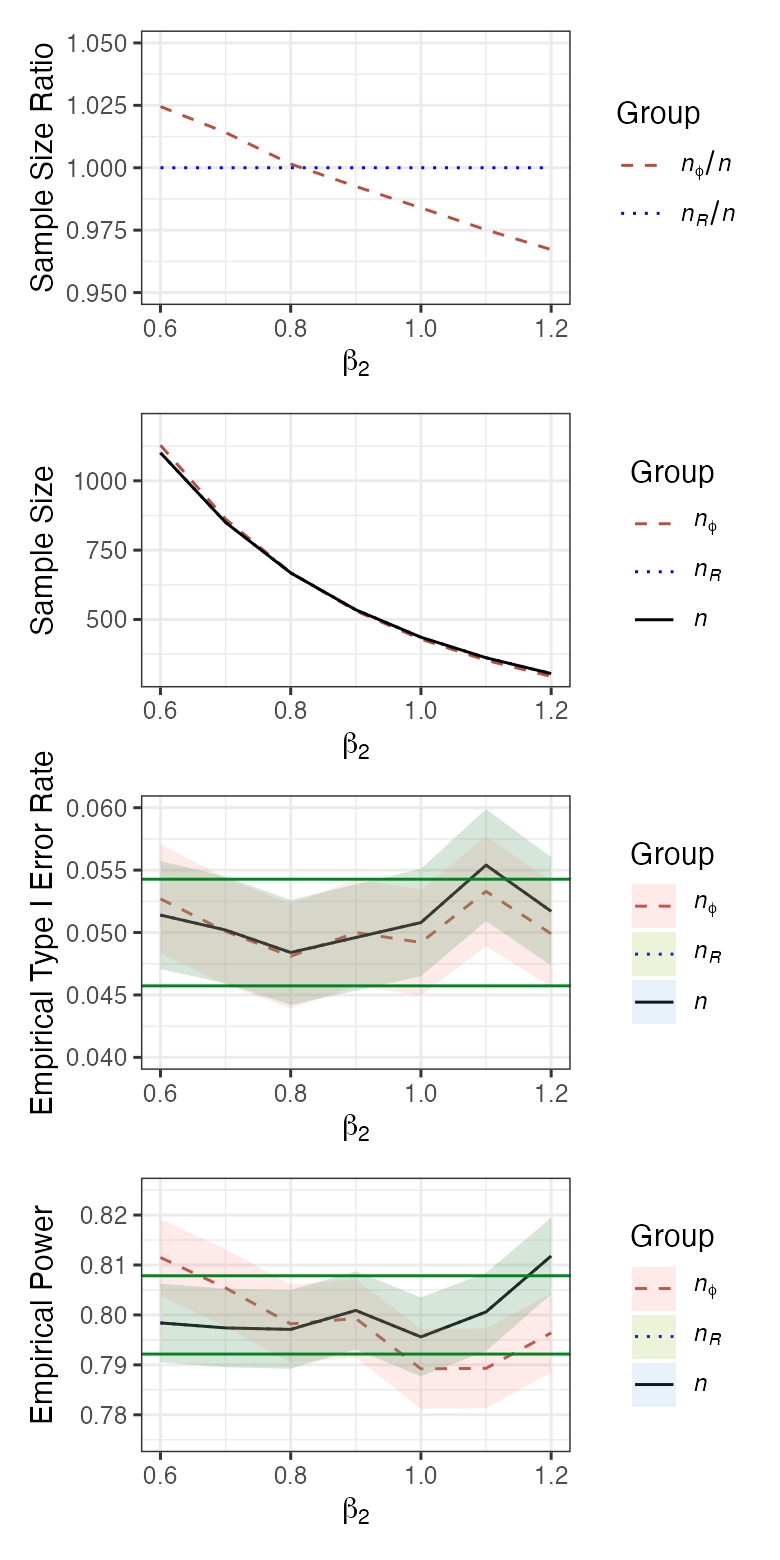}
    \caption{var $\propto$ mean}
    \end{subfigure}
    \begin{subfigure}{.32\textwidth}
    \centering
    \includegraphics[width=\linewidth]{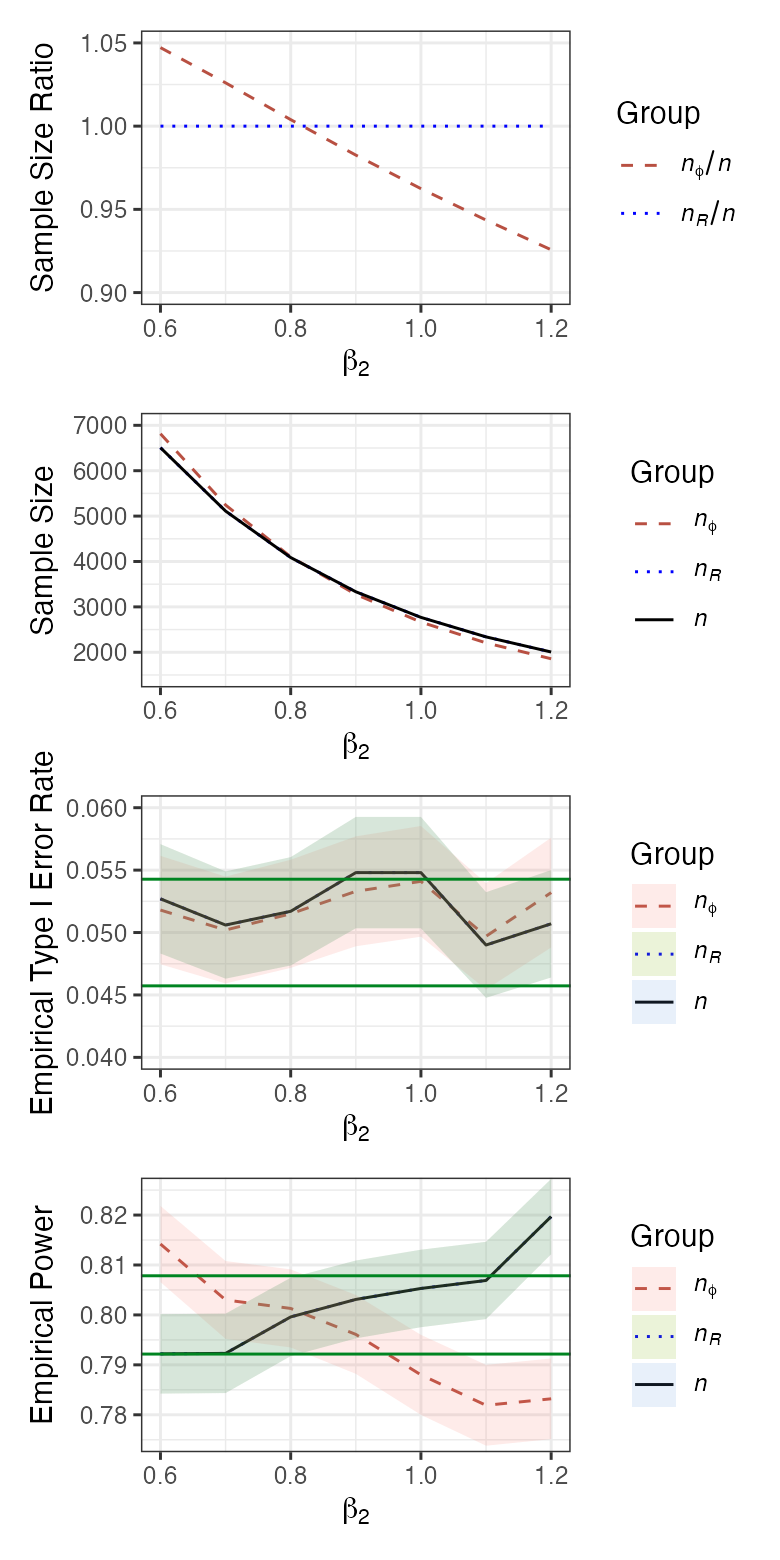}
    \caption{var $\propto$ mean$^2$}
    \end{subfigure}
    \caption{The empirical type I error rates and power of the Wald test based on the simulated count data for three cases of outcome variables, using a identity link and  sample sizes $n$, $n_{\phi}$, and $n_R$. The X-axis represents the coefficient $\beta_2$. The shaded envelopes represent the 95\% confidence intervals of the empirical type I error rate and empirical power, respectively. The two horizontal lines in each panel indicate the 95\% confidence interval around the corresponding target level (0.05 for type I error and 0.8 for power), serving as a benchmark for good calibration. Results falling within these horizontal lines indicate well-calibrated performance.}\label{fig:count_identity_vary_beta}
\end{figure}

Under the identity link, we set $\lambda_1 = 4$, $\lambda_2 = 0.4$, and $\beta_1 = 0.4$, and allowed $\beta_2$ to range from 0.6 to 1.2. As with the log link, the same three count outcome distributions were examined. This setting provides a complementary view where the link function is linear, favoring the performance of the P2R2 approximation. We again compared $n$, $n_\phi$, and $n_R$ and their associated empirical powers across the range of $\beta_2$ values.
Figure~\ref{fig:count_identity_vary_beta} shows the corresponding results for the identity link.

\subsection{Positive Continuous Outcomes} 
\begin{figure}[!htb] 
    \centering
    \begin{subfigure}{.36\textwidth}
    \centering
    \includegraphics[width=\linewidth]{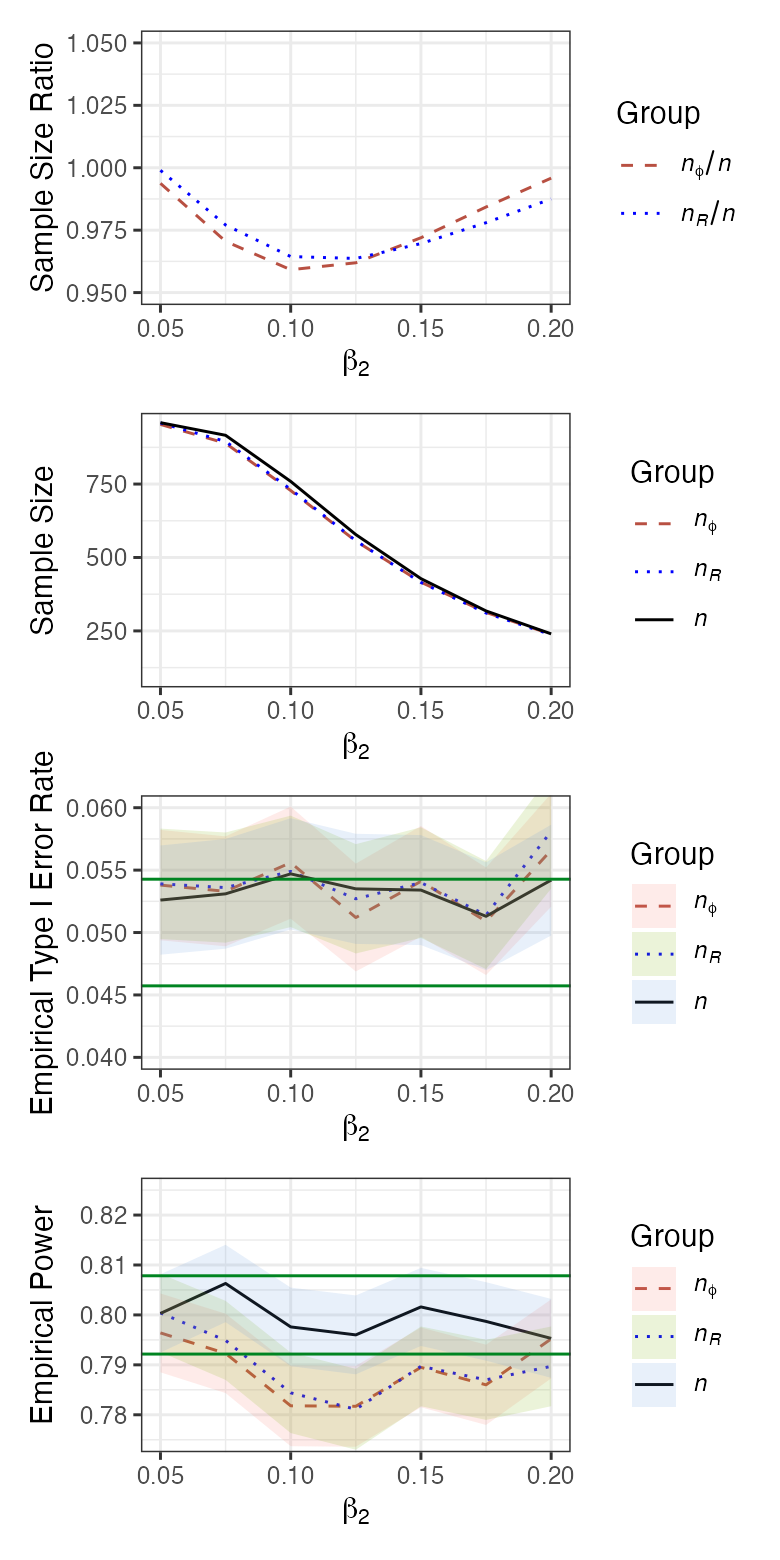}
    \caption{var $\propto$ mean} \label{fig:positive_log_1_vary_beta}
    \end{subfigure}
    \begin{subfigure}{.36\textwidth}
    \centering
    \includegraphics[width=\linewidth]{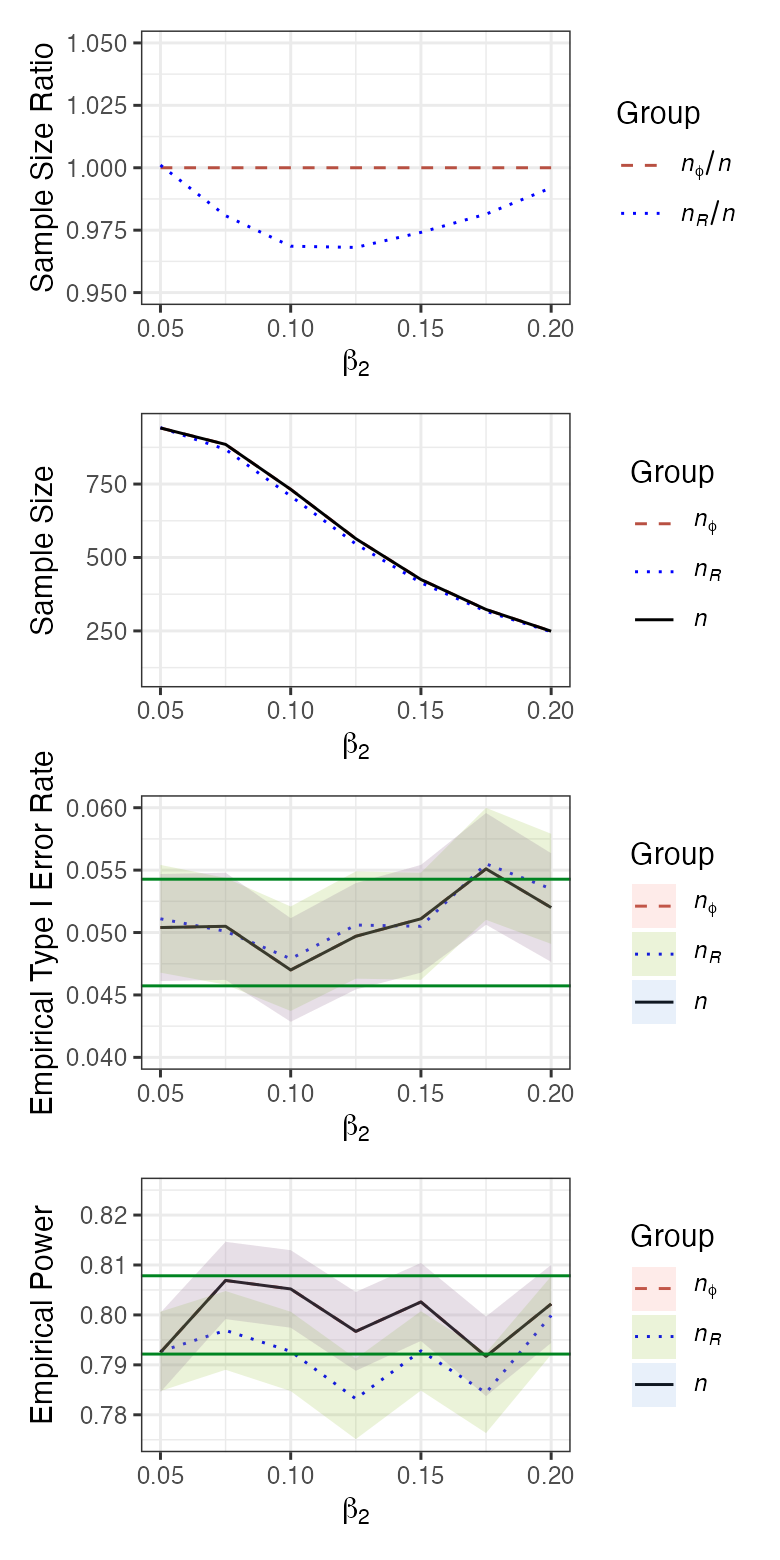}
    \caption{var $\propto$ mean$^2$}\label{fig:positive_log_2_vary_beta}
    \end{subfigure}
    \caption{The empirical type I error rates and power of the Wald test based on the simulated positive continuous data for two cases of outcome variables, using a log link and  sample sizes $n$, $n_{\phi}$, and $n_R$. The X-axis represents the coefficient $\beta_2$. The shaded envelopes represent the 95\% confidence intervals of the empirical type I error rate and empirical power, respectively. The two horizontal lines in each panel indicate the 95\% confidence interval around the corresponding target level (0.05 for type I error and 0.8 for power), serving as a benchmark for good calibration. Results falling within these horizontal lines indicate well-calibrated performance.}\label{fig:positive_log_vary_beta}
\end{figure}

For positive continuous outcomes, we considered gamma distributions with variance proportional to the mean and to the squared mean, using the same log link function as in Section~\ref{sec:pos_cont_outcome}. Coefficients were set to $\lambda_1 = 1$, $\lambda_2 = 0.15$, and $\beta_1 = 0.1$, with $\beta_2$ ranging from 0.05 to 0.2. 
Figure~\ref{fig:positive_log_vary_beta} presents the results for the continuous-outcome settings.

\end{appendices}

\end{document}